\documentclass[prb,twocolumn,showpacs,preprintnumbers,amsmath,amssymb]{revtex4}

\usepackage{graphicx}
\usepackage{dcolumn}
\usepackage{bm}

\newcommand{\Z}{{\Bbb Z}}                                    
\newcommand{\C}{{\Bbb C}}                                    
\newcommand{\Cplus}{{\Bbb C}^+}                              
\newcommand{\cf}[1]{\langle #1 \rangle}                      
\newcommand{\bra}[1]{\langle #1 \!\mid\!}                    
\newcommand{\ket}[1]{\!\mid\! #1 \rangle}                    
\newcommand{\no}[1]{: \! #1 \! :}                            
\newcommand{\im}{\mbox{Im} \,}                               
\newcommand{\onehalf}{\mbox{$\frac{1}{2}$}}                  
\newcommand{\1}{\openone}                                    
\newcommand{\vJ}{\mbox{\boldmath $J$}}                       
\newcommand{\vphi}{\mbox{\boldmath $\phi$}}                  
\newcommand{\phdagger}{\mathop{\phantom{\dagger}}}           
\newcommand{\psiop}[1]{\psi^{\phdagger}_{#1}}                
\newcommand{\psidop}[1]{\psi^{\dagger}_{#1}}                 
\newcommand{\bop}[1]{b^{\phdagger}_{#1}}                     
\newcommand{\bdop}[1]{b^{\dagger}_{#1}}                      
\newcommand{\fop}[1]{f^{\phdagger}_{#1}}                     
\newcommand{\fdop}[1]{f^{\dagger}_{#1}}                      

\newcommand{\del}{\partial}
\newcommand{\bml}{\begin{mathletters}}                           
\newcommand{\eml}{\end{mathletters} \hspace{-5pt}}       

\begin{document}

\title{Critical Theory of the Two-Channel Anderson Impurity Model} 

\author{Henrik~Johannesson}

\affiliation{Institute of Theoretical Physics, 
Chalmers University of Technology and G\"oteborg University, \\
SE-412 96 G\"oteborg, Sweden}

\author{N.~Andrei}

\author{C.~J.~Bolech}
 \altaffiliation[Present address at ]{Universit\'e de Gen\`eve, Switzerland.}

\affiliation{Center for Materials Theory, Serin Physics Laboratory, 
Rutgers University \\ Piscataway, New Jersey 08854-8019}


\begin{abstract}
  We construct the boundary conformal field theory that describes the
  low-temperature behavior of the two-channel Anderson impurity
  model. The presence of an exactly marginal operator is shown to
  generate a line of stable fixed points parameterized by the charge
  valence $n_c$ of the impurity.  We calculate the exact
  zero-temperature entropy and impurity thermodynamics along the fixed
  line.  We also derive the critical exponents of the characteristic
  Fermi edge singularities caused by time-dependent hybridization
  between conduction electrons and impurity.  Our results suggest that
  in the mixed-valent regime ($n_c \neq 0, 1$) the electrons
  participate in two competing processes, leading to frustrated
  screening of spin {\em and} channel degrees of freedom. By combining
  the boundary conformal field theory with  the Bethe
  Ansatz solution we obtain a complete description of the low-energy
  dynamics of the model.
\end{abstract}

\pacs{71.27.+a, 75.20.Hr, 75.40.-s}
\keywords{Kondo effect, Heavy Fermions, Non-Fermi Liquid}

\maketitle

\newpage


\section{Introduction}
\label{section1}

In recent years, a growing class of materials has been shown to exhibit
metallic behaviors that violate Landau's Fermi liquid theory.
\cite{LandauReview} Examples include the ``normal phases'' of underdoped and
optimally doped high-$T_c$ cuprates,\cite{CuprateReview} a variety of (quasi)
one-dimensional conductors -- ranging from single-wall carbon nanotubes
\cite{NanoReview} to the Bechgaard salts\cite{BechgardReview} -- several
artificially designed nanostructures,\cite{DesignReview} as well as certain
cerium- and uranium-based heavy fermion alloys.\cite{HeavyFermionReview} In
light of the spectacular success of Landau's theory in explaining the
properties of conventional metals, the proliferation of experimental systems
that depart from its predictions presents a challenge to the theorist. Not
forming a Fermi liquid, the mobile electrons of these systems cannot be
adiabatically connected to a noninteracting electron gas, and their
description requires a theoretical approach of a different brand.

A particularly intriguing case is that of the heavy fermion compound
UBe$_{13}$. Like several other metallic materials containing rare earth or
actinide ions with partially filled f-electron shells UBe$_{13}$ exhibits
manifest non-Fermi liquid behavior at low temperatures. In particular,
the specific heat shows a $T\ln T$ behavior all the way
down to the superconducting transition (at roughly 1K),\cite{UBeExp1}
suggestive of a two-channel overscreened Kondo effect driven by the
f-electrons of the uranium ions. There are two possible integer valence
configurations for uranium embedded in a Be$_{13}$ host: a $\Gamma_6$ magnetic
spin doublet in the $5f^3$ configuration and a $\Gamma_3$ electrical
quadrupolar doublet in the $5f^2$ configuration, the two levels being
separated by an energy $ \varepsilon$. Arguing that the extremely
weak magnetic field dependence of the material \cite{HeavyFermionReview}
excludes the usual magnetic Kondo effect, Cox \cite{Cox} proposed that the
observed anomalies instead derive from a quenching of the quadrupolar degrees
of freedom by the local orbital motion of the $\Gamma_8$ conduction electrons
in the material (with the electron spin $\sigma = \, \uparrow, \downarrow$
providing the two ``channels'' required for overscreening). The proposal has
been controversial,\cite{AndrakaTsvelik} however, and experimental data on
the nonlinear magnetic susceptibility indicate that the low-lying magnetic
excitations are rather predominantly dipolar in character.\cite{Ramirez}
Whether the energy splitting $ \varepsilon$ between the two
doublets is sufficiently large for a quadrupolar Kondo scenario to become
viable is an additional unresolved issue. Aliev {\em et al.}\cite{Aliev1}
have suggested that the magnetic and quadrupolar doublets may in fact be
near degenerate, leading instead to a mixed-valent state with a novel type of
interplay between quadrupolar and magnetic Kondo-type screening. Indirect
support for this hypothesis comes from non-linear susceptibility\cite{Aliev1}
and other\cite{Aliev2} measurements on the thoriated compound
U$_{0.9}$Th$_{0.1}$Be$_{13}$. The presence of thorium ions suppresses the
superconducting transition in the undoped material and one observes instead a
crossover from a magnetic to a quadrupolar groundstate at low temperatures.

The simplest model that captures a possible mixed-valent regime with both
magnetic and quadrupolar character is the two-channel Anderson single-impurity
model.\cite{CoxZawadowski,SchillerAndersCox,KrohaWolfle} In this model the 
conduction electrons carry both
spin $(\sigma = \uparrow, \downarrow)$ and quadrupolar $(\alpha = \pm)$
quantum numbers, and hybridize via a matrix element $V$ with a local uranium
ion. The ion is modeled by a quadrupolar ($5f^2$) doublet created by a boson
operator $\bdop{\alpha}$ and a magnetic ($5f^3$) doublet created by a fermion
operator $\fdop{\sigma}$. Strong Coulomb repulsion implies that the localized
$f$-levels can carry at most one electron, and this condition is implemented
as an operator identity for the pseudoparticles: $ \fdop{\sigma} \fop{\sigma}
+ \bdop{\bar{\alpha}} \bop{\bar{\alpha}} = 1$, where $\bar{\alpha}$ denotes
the conjugate representation.  With these provisos the Hamiltonian is written
as
\begin{equation}   \label{hamiltonian}
H =  H_{bulk} + H_{ion} + H_{hybr} \, ,  
\end{equation}
where
\begin{eqnarray}  
H_{bulk} & = & \int dx \no{\psidop{\alpha \sigma}(x) (-i\partial_x)
\psiop{\alpha \sigma}(x)}
\label{bulk} \\
H_{ion} & = & \varepsilon_s \fdop{\sigma} \fop{\sigma} + 
\varepsilon_q \bdop{\bar{\alpha}} \bop{\bar{\alpha}}
\label{ion} \\
H_{hybr} & = & V [\psidop{\alpha\sigma}(0) \bdop{\bar{\alpha}} \fop{\sigma} +  
\fdop{\sigma} \bop{\bar{\alpha}} \psiop{\alpha\sigma}(0)] \, .
\label{hybr}
\end{eqnarray}
Here the conduction electrons are described by one-dimensional fields
$\psidop{\alpha \sigma}(x)$, matching the $\Gamma_8$ representation
(appearing in the reduction of $\Gamma_3 \times \Gamma_6$ and thus
coupling to the impurity). The electron fields are chiral, obtained -
as usual\cite{AndreiFuruyaLowenstein} - by reflecting the outgoing
radial $(x > 0)$, $\Gamma_8$-component of the original three-dimensional
problem to the negative $x$-axis and disregarding components that do
not couple to the ion at $x=0$.  The spectrum is linearized around the
Fermi level and the Fermi velocity is set to unity, with the resulting
density of states being $\rho = 1/(2\pi)$. The normal ordering is
taken with respect to the filled Fermi sea, and the energies $\varepsilon_s$ and
$\varepsilon_q$ are those of the magnetic and quadrupolar doublets
respectively.

Recently two of us presented an exact solution of the model, based on
a Bethe Ansatz construction.\cite{BolechAndrei} A complete
determination of the energy spectrum and the thermodynamics was given,
allowing a full description of the evolution of the impurity from its
high temperature behavior with all four impurity states being equally
populated down to the low energy dynamics characterized by a line of
fixed point Hamiltonians $H^*(\varepsilon, \Gamma)$ where $\varepsilon
=\varepsilon _{s}-\varepsilon _{q}$ and $\Gamma=\pi\rho V^2$ ($\Gamma$
is held fixed in what follows).

It is found that the line of fixed points is characterized by a
zero-temperature entropy $S^0_{\text{imp}}=k_B\ln \sqrt{2}$ and a specific
heat $C_{\text{v}}^{\text{imp}}\sim T\ln T$ typical of the two-channel Kondo
fixed point. However, the physics along the line varies with $\varepsilon$.
Consider $n_{c}=\left\langle f_{\sigma }^{\dagger }f_{\sigma }\right\rangle $,
the amount of charge localized at the impurity. For
$\varepsilon\alt\mu-\Gamma$, one finds that $n_c\approx 1$, signaling the
magnetic integral valence regime. At intermediate temperatures a magnetic
moment forms which undergoes frustrated screening as the temperature is
lowered, leading to zero-temperature anomalous entropy and anomalous specific
heat. For $\varepsilon\agt\mu+\Gamma$ the system is in the quadrupolar integral 
regime and a quadrupolar moment forms. In the mixed valence regime,
$|\varepsilon-\mu|\alt\Gamma$, similar low-temperature behavior is observed
though without the intermediate regime of moment formation. In more detail,
each point on the line of fixed points is characterized by two energy scales
$T_{l,h}(\varepsilon)$. These scales describe the quenching of the entropy as
the temperature is lowered: the first stage taking place at the high-
temperature scale $T_{h}$, quenching the entropy from $k_B\ln 4$ to $k_B\ln
2$, the second stage at $T_{l}$, quenching it from $k_B\ln 2$ to $k_B\ln
\sqrt{2}$. In the integral valence regime, $|\varepsilon-\mu|\gg\Gamma$, the
two scales are well separated and as long as the temperature falls between
these values a moment is present (magnetic or quadrupolar depending on the
sign of $\varepsilon-\mu$), manifested by a finite temperature plateau
$S_{\text{imp}}=k_B\ln2$ in the entropy. It is quenched when the temperature
is lowered below $T_{l}$, with $T_{l} \to T_{K}$ in this regime. For
$\varepsilon=\mu$ the scales are equal, $T_{l}(\mu)=T_{h}(\mu)$, and the
quenching occurs in a single stage.

In the present paper we shall concentrate on the low-energy regime and
give a detailed description of the line of fixed points in terms of
boundary conformal field theory $-$ the low-energy effective
Hamiltonian.  It was argued by Affleck and Ludwig
\cite{Affleck,AffleckLudwig1} that the low-energy regime of a quantum
impurity problem is described by a boundary conformal field theory
(BCFT) with a conformally invariant boundary condition replacing the
dynamical impurity. However, to identify the effective low-energy
theory characterizing a given microscopic impurity model, such as
Hamiltonian (\ref{hamiltonian}), it is necessary to employ more
powerful methods, valid over the full energy range, that can connect the
microscopic Hamiltonian  to the
effective low-energy theory.  One needs in principle to carry out a
full renormalization group calculation or, when available, use an
exact solution to extract the BCFT.

Although applicable only in the neighborhood of a low-temperature fixed point,
a BCFT formulation has certain advantages. First, it provides an elegant
scheme in which to analyze the approach to criticality.  The leading scaling
operators can be identified explicitly, yielding access to the low-temperature
thermodynamics in closed analytical form. These can be matched with the
expressions from the Bethe-Ansatz solution allowing the determination of the
parameters appearing in the BCFT. This will be carried out below. Secondly $-$
and more importantly $-$ having identified the scaling operators, the asymptotic
dynamical properties (Green's functions, resistivities, optical
conductivities, etc.)  can in principle be calculated. These results 
will be presented in a subsequent work.

The paper is organized as follows. In Sec.~II we review the basics of BCFT,
with particular focus on applications to quantum impurity problems. In
Sec.~III we derive the specific BCFT which describes the low-temperature
physics of the two-channel Anderson model in Eq. (\ref{hamiltonian}). Some
important features of this BCFT are discussed, and in this section we also
apply it to the Fermi edge singularity problem for this model. In Sec.~IV we
then combine the results obtained with the Bethe Ansatz solution
\cite{BolechAndrei} to analytically calculate the zero-temperature impurity
entropy, the impurity contributions to the low-temperature specific heat,
and the linear and non-linear impurity magnetic susceptibilities. Section ~V, finally,
contains a summary and a discussion of our results.

\section{Boundary Conformal Field Theory Approach}
\label{section2}

The BCFT approach to quantum impurity problems is well reviewed in the
literature,\cite{AffleckReview,LudwigReview,SaleurReview} and we here only
collect some basic results so as to fix conventions and notation.

The key idea is to trade a local impurity-electron interaction for a scale
invariant boundary condition on the critical theory that represents the
extended (electron) degrees of freedom. In the presence of a boundary, the
left- and right-moving parts of the fields $\varphi(t,x)$ in the
critical theory are identified via analytic continuation beyond the boundary
$x=0$, such that
\begin{equation}    \label{Nonlocal}
\varphi(t,x) = \varphi_L(t,x)\varphi_R(t,x) \ \rightarrow \ 
\varphi_L(t,x)\varphi_L(t,-x) .
\end{equation} 
Thus, the boundary effectively turns the fields nonlocal. The consequence of
this can be delineated via the operator product expansion
\begin{multline}      \label{BoundaryOPE}
\varphi_L(z) \varphi_L(z^{\ast}) \rightarrow \sum_j \frac{g_j}
{ (z-z^{\ast})^{2\Delta - \Delta_j}}
{\cal O}^{(j)}_L (0, \tau) \, , \\
z \rightarrow z^{\ast} \, .
\end{multline}
We have here introduced a complex variable $z = \tau + ix$, with $\tau$
the Euclidean time and $x$ the spatial coordinate. The function $g_j \ ( = 0$ or
$1)$ selects the {\em boundary operators} ${\cal O}^{(j)}_L$
associated with the given boundary condition, $\Delta_j \ (\Delta)$ being the
dimension of ${\cal O}^{(j)}_L \ ( \varphi_L )$. It follows from
Eqs. (\ref{Nonlocal}) and (\ref{BoundaryOPE}) that an $n$-point function of a field
$\varphi$ close to the boundary turns into a 
linear combination of $n$-point functions of the {\em
chiral} boundary operators ${\cal O}^{(j)}_L$.  In particular, this implies
that the {\em boundary scaling dimension} $\Delta_{bound}$ of $\varphi$ that
governs its autocorrelation function close to the boundary,
\begin{multline}    \label{Auto}
\langle \varphi(t,x) \varphi(0,x) \rangle - \langle \varphi(t,x) \rangle
\langle \varphi(0,x) \rangle \ \sim \ t^{-2\Delta_{bound}} \  \\
 |t| >> |x| 
\end{multline}
is precisely given by the scaling dimension of the {\em leading} boundary
operator appearing in Eq. (\ref{BoundaryOPE}).

This sets the strategy for treating a quantum impurity problem: identify first
the particular boundary condition that plays the role of the impurity
interaction. Given that this boundary condition is indeed scale invariant
(together with the original bulk theory), one can then use the machinery of
BCFT to extract the corresponding boundary scaling dimensions. As these
determine the asymptotic autocorrelation functions [cf. Eq. (\ref{Auto})] the
finite-temperature properties due to the presence of the boundary ({\em alias}
the impurity) are easily accessed from standard finite-size scaling by
treating the (Euclidean) time as an inverse temperature.

To identify the ``right'' boundary condition it is convenient to exploit a
well-known result\cite{CardyFinite} from conformal field theory, relating the
energy levels in a finite geometry to the (boundary) scaling dimensions of
operators in the (semi-) infinite plane. More precisely, consider a
conformally invariant theory defined on the strip $\{ w= u + iv \ | \ -\infty
< u < \infty ,\ 0 \leqslant v \leqslant \ell \}$ with $u$ the Euclidean time and
$v$ the space coordinate. Then impose a conformally invariant boundary
condition, call it $A$, at the edges $v=0$ and $v=\ell $, and map the strip
onto the semi-infinite plane $\{z = {\tau} + ix | x \geqslant 0 \}$ using the
conformal transformation $z = \exp (\pi w/l)$ (implying boundary condition $A$
at $x =0$). With $E_0$ the ground-state energy, one has
\begin{equation}     \label{CardyFormula}
E_n = E_0 + \frac{\pi  \Delta_n } {\ell},
\end{equation}
where $\{ E_n \}$ is the spectrum of excited energy levels in $0 \leqslant v
\leqslant \ell$ and $ \{ \Delta_n \} $ is the spectrum of {\em boundary
scaling dimensions} in the semi-infinite plane.

The problem is thus reduced to determining the finite-size spectrum of the
theory. In certain privileged cases this can be done by direct calculation.
Alternatively, one hypothesizes a boundary condition and compares the
consequences with known answers. An example is the {\em fusion hypothesis} of
Affleck and Ludwig.\cite{AffleckLudwig1} One here starts with some known,
trivial boundary condition on the critical theory, with no coupling to the
impurity.  For free electrons carrying charge, spin (and maybe some
additional ``flavor'' degrees of freedom) the spectrum organizes into a
conformal tower structure,\cite{DiFrancesco} with the charge, spin and flavor
sectors ``glued'' together so as to correctly represent the electrons {\em
  (Fermi liquid gluing condition)}. Then, turning on the electron-impurity
interaction, its only effect $-$ according to the fusion hypothesis $-$ is 
to replace this gluing condition by some new nontrivial gluing of the
conformal towers. In the case of the $m$-channel Kondo problem, the new gluing
condition is obtained via {\em Kac-Moody fusion}\cite{ItzyksonDrouffe} with
the spin-1/2 primary operator that ``corresponds'' to the Kondo impurity. As a
consequence, the conformal tower with spin quantum number $j_s$ is mapped onto
new towers labeled by $j^{\prime}_s$, where $j^{\prime}_s = |j_s-\onehalf|,
|j_s-\onehalf| + 1, ..., \mbox{min} \{j_s+\onehalf,m-j_s-\onehalf\}.$ This is
the essence of the BCFT approach, as applied to the Kondo problem.

For the two-channel Anderson model in Eq. (\ref{hamiltonian}) it is {\em a priori}
less obvious which operator to use for fusion. Here we shall instead identify
the finite-size spectrum by studying certain known limiting cases, guided by
the exact Bethe Ansatz solution of the model.\cite{BolechAndrei} As it turns
out, the spectrum thus obtained can be reconstructed by fusion with the
leading {\em flavor} boundary operator (representing the channel, or {\em
quadrupolar}, degrees of freedom), but {\em in addition there occurs an
effective renormalization of the charge sector}. This signals the novel
aspect of the present problem. A technical remark may here be appropriate:
Since the boundary scaling dimensions in Eq. (\ref{CardyFormula}) are connected to
energy levels in a strip with the {\em same} boundary condition at the two
edges, we are
effectively considering a finite-size energy spectrum with {\em two} quantum
impurities present, one at each edge of the strip. Formally, this can be taken
care of by performing Kac-Moody fusion {\em twice} with the relevant primary
operator {\em (double fusion)}. This is an important point, not always fully
appreciated in the literature.

Let us finally mention that there exists another, more fundamental description
of BCFT,\cite{Cardy} based on the notion of {\em boundary states},
particularly useful when studying zero-temperature properties of a quantum
impurity problem. We shall give a brief exposition of it in Sec.~(IV.A), where
we use it for analyzing the zero-temperature entropy contributed by the
impurity.

\section{Finite-Size Spectrum and Scaling Operators}
\label{section3}

As we discussed in the previous section, the scaling dimensions of the
boundary operators that govern the critical behavior are in one-to-one
correspondence with the levels of the finite-size spectrum of the model. In
the present case the spectrum can in principle be constructed from the exact
Bethe Ansatz solution in Ref.~\onlinecite{BolechAndrei}, but this requires an
elaborate
analysis. The reason is that the solution takes the
form of a so called ``string solution'' even in the groundstate. Since a
string solution is valid only in the thermodynamic limit, an application of
standard finite-size techniques\cite{deVegaWoynarovich} would lead to
incorrect results.  One may instead pursue another strategy and identify the
finite size spectrum (hence the dimensions of the scaling operators) that
reproduces the results of the exact solution in the low-energy limit.  In our
case the search for the wanted spectrum can be carried out efficiently by
combining symmetry arguments with known results for the finite-size spectra
for two limiting cases of the Hamiltonian (\ref{hamiltonian}): the ordinary
Anderson model and the two-channel Kondo model.  We shall see that the low
energy spectrum thus obtained, Eq.~(\ref{mChannelAndersonSpectrum}) below,
leads indeed to a line of fixed points characterized by entropy $S=$ k$_B \ln
\sqrt{2}$ and susceptibility $C_V \, \sim T \ln T $ as required by the Bethe
Ansatz solution. The fitting of the proposed spectrum to the solution also
allows the determination of all thermodynamically relevant parameters in the BCFT.

\subsection{Identifying the Critical Theory}

Let us begin by reviewing the single channel Anderson and Kondo
models. The latter is obtained in the integer-valence limit $n_c
\rightarrow 1$ (with $n_c \equiv \cf{\fdop{\sigma} \fop{\sigma}}$
measuring the charge localized at the impurity site) where a magnetic
moment forms, signaling the entrance to the Kondo regime.\cite{Hewson}
It is instructive to consider this case in some detail before
proceeding to an analysis of the full model in Eq. (\ref{hamiltonian}).

As is well known, the ordinary Anderson model\cite{Anderson} with integer
valence
$n_c = 1$ can be mapped onto the single-channel Kondo model via a
Schrieffer-Wolff transformation.\cite{Hewson} To $O(1/{\ell})$, with
${\ell}$ the length of the system, the finite-size energy spectrum of
the Kondo model takes the form \cite{AffleckLudwig1}
\begin{equation}    \label{KondoSpectrum}
E = E_0 + \frac{\pi}{{\ell}} \left[\left( \, \frac{1}{4}Q^2 + N_c \, \right) +
\left(\, \frac{j_s(j_s+1)}{3} + N_s\, \right) \right] \, ,
\end{equation}
where the Fermi velocity has been set to unity, as in Eq. (\ref{hamiltonian}).
Here $Q \in \Z$ and $j_s =0, 1/2$ are charge and spin quantum numbers --
defined with respect to the filled Fermi sea -- and labeling the corresponding Kac-Moody
{\em primary states}.\cite{ItzyksonDrouffe} The positive integers $N_c$ and
$N_s$ index the {\em descendant levels} in the associated U(1) and SU(2)$_1$
conformal towers.  The quantum numbers are constrained to appear in the
combinations
\begin{eqnarray}   \label{KondoGluing}
Q = 0 \ \mbox{mod} \  2, \ &  &\  j_s = \frac{1}{2}  \nonumber \\
Q = 1 \ \mbox{mod} \  2,  \ &  & \ j_s = 0
\end{eqnarray}
which define the {\em Kondo gluing conditions} for charge and spin conformal
towers.

We can rewrite this result in a form immediately generalizable to the Anderson
model, viewing it as the $n_c \rightarrow 1$ limit of the Anderson model
spectrum.  
Redefine the charge
quantum number in Eqs. (\ref{KondoSpectrum}) and (\ref{KondoGluing}): $Q
\rightarrow Q - 1$. The offset mirrors the fact that the Anderson impurity
carries charge (in contrast to a Kondo impurity) and, therefore, shifts the net
amount of charge in the groundstate compared to the Kondo case. We thus write
for the finite-size spectrum of the Anderson model in the $n_c \rightarrow 1$
limit:
\begin{widetext}
\begin{equation} \label{AndersonSpectrum1}
E = E_0 +  \frac{\pi}{{\ell}} \left[\left(\frac{1}{4}(Q- 1)^2 
- \frac{1}{4}+ N_c \right) 
 +  \left(\frac{j_s(j_s+1)}{3} + N_s \right)\right] \, ,
\end{equation}
\end{widetext}
with
\begin{eqnarray}   \label{AndersonGluing}
Q = 0 \ \mbox{mod}\ 2,  \ &  &\  j_s = 0 \nonumber \\
Q = 1 \ \mbox{mod}\ 2, \ &  &\  j_s =\frac{1}{2}
\end{eqnarray}
Note that the constraints in Eqs. (\ref{AndersonGluing}) are the same as those for
the charge and spin conformal towers of free electrons {\em (Fermi liquid
gluing conditions)}. Also note that we have renormalized the groundstate
energy, so that $E_0 = E(Q=0)$, by subtracting off a constant $\pi/4{\ell}$.
The redefinition of charge quantum numbers implies a relabeling of levels: 
the $Q$-level in Eq. (\ref{KondoSpectrum}) (constrained by Eqs. (\ref{KondoGluing}))
corresponds to the $(Q-1)$-level
in Eq. (\ref{AndersonSpectrum1}) (with the constraint (\ref{AndersonGluing})), 
modulo the $\pi/4{\ell}$ shift.

Let us now consider the spectrum of the Anderson model away from the Kondo
limit,
$n_c \neq 1$. Assuming the absence of any relevant operator
[in the renormalization group (RG) sense] that could take the model to a 
new boundary fixed point --~and
hence change the gluing conditions in Eqs. (\ref{AndersonGluing})~-- we expect that
the finite-size spectrum still has a Kac-Moody structure with Fermi liquid
gluing conditions. Since the groundstate now has an (average) surplus charge
$n_c \neq 1$ compared to the case with an empty localized level, we conclude
from Eq. (\ref{KondoSpectrum}) that
\begin{widetext}
\begin{equation} \label{AndersonSpectrum2}
E = E_0 + \frac{\pi}{{\ell}} \left[ \left(\, \frac{1}{4}(Q-n_c)^2 - 
(\frac{n_c}{2})^2 + N_c \right) +
\left( \frac{j_s(j_s+1)}{3} + N_s \right)\right] \, , 
\end{equation}
\end{widetext}
with $Q$ and $j_s$ constrained by Eqs. (\ref{AndersonGluing}).  Note that this
result has been extracted from the Kondo finite-size spectrum in
Eqs. (\ref{KondoSpectrum}) and (\ref{KondoGluing}) simply by keeping track on the
offset of charge with respect to the filled Fermi sea as the impurity acquires a
nonzero charge valence $n_c$. The additional assumption that no relevant
operator is generated as we move away from the Kondo limit $n_c = 1$ (and that
therefore the gluing conditions in Eqs. (\ref{KondoGluing}) remain unchanged), will
be shown to be self-consistent when we analyze the full two-channel Anderson
model.

We can corroborate our procedure by comparing our result in
Eq. (\ref{AndersonSpectrum2}) with that obtained by Fujimoto and collaborators.
\cite{FujimotoKawakamiYang2} Applying standard finite-size techniques
\cite{deVegaWoynarovich} to the exact Bethe Ansatz solution\cite{Wiegmann} of
the Anderson model these authors found that
\begin{widetext}
\begin{equation} \label{FujimotoSpectrum}
E = E_0 + \frac{\pi}{{\ell}} \left[\left(\,
\frac{1}{4}(Q-\frac{2\delta_F}{\pi})^2 -
(\frac{\delta_F}{\pi})^2 + N_c
\, \right) +
\left(\, \frac{j_s(j_s+1)}{3} + N_s\, \right)\right] + O(\frac{1}{{\ell}^2}) \,
,
\end{equation}
\end{widetext}
with Fermi liquid gluing (\ref{AndersonGluing}) of $Q$ and $j_s$.  The phase
shift $\delta_F \equiv \delta_{\sigma}({\epsilon_F})$ appearing in
(\ref{FujimotoSpectrum}) is that of an electron with spin $\sigma$ at the
Fermi level, related  to the average charge $n_c \equiv \cf{\fdop{\sigma}
\fop{\sigma}}$ at the impurity site by the Friedel-Langreth sum rule
\cite{Langreth}
\begin{equation}   \label{FriedelLangreth}
\delta_F = \frac{\pi}{2} n_c \, .
\end{equation}
Inserting Eq. (\ref{FriedelLangreth}) into Eq. (\ref{FujimotoSpectrum}), we
immediately recover our result (\ref{AndersonSpectrum2}).

Before returning to the full two-channel Anderson model, let us point out that
the duality between the charge shift $Q \leftrightarrow Q -1$ in the integer
valence limit $n_c=1$ and the change of gluing conditions
(\ref{AndersonGluing}) $\leftrightarrow$ (\ref{KondoGluing}) simply reflects
the well-known fact that the ordinary Kondo effect can formally be described
as a local potential scattering of otherwise free electrons, causing a phase
shift $\delta_F = \pi/2$ of their wave functions
[cf.~Eq.~(\ref{FriedelLangreth})]. It is instructive to make this duality
transparent by identifying the effective low-energy Hamiltonian $H_{charge}$
that produces the charge spectrum in Eq. (\ref{AndersonSpectrum1}). Introducing
the charge currents
\begin{equation}     \label{ChargeCurrent}
J(x) = \no{\psidop{\alpha \sigma}(x) \psiop{\alpha \sigma}(x)}
\end{equation}
(where as before the normal ordering is taken with respect to the filled Fermi sea),
one immediately recognizes the charge part of (\ref{AndersonSpectrum1}) as the
spectrum of the Sugawara Hamiltonian \cite{ItzyksonDrouffe}
\begin{equation}   \label{SugawaraCharge}
H_{charge} = \frac{1}{4} \int dx \no{J(x) J(x)}  - \, \frac{1}{2} J(0) \, .
\end{equation}
The term $\sim J(0)$ explicitly reveals the presence of an effective local
scattering potential. By redefining the charge quantum numbers in the integer
valence limit, $Q - 1 \rightarrow Q$, this potential is disguised as a
renormalized boundary condition, encoded in the Kondo gluing conditions
(\ref{KondoGluing}).

With these preliminaries let us now go back to the full two-channel Anderson
model in Eq. (\ref{hamiltonian}), first considering the case when $ \varepsilon =
\varepsilon_s - \varepsilon_q \ll \mu - \Gamma$. From the Bethe Ansatz
solution\cite{BolechAndrei} one finds that in this limit the physics is that
of the overscreened two-channel (magnetic) Kondo model.\cite{NozieresBlandin}  
This is expected,
since for this case the ion has integer valence, with an associated magnetic
moment of spin 1/2. The finite-size spectrum of the two-channel Kondo model
has been derived from that of free electrons (carrying spin {\em and} flavor)
via {\em Kac-Moody fusion} \cite{ItzyksonDrouffe} with the spin-1/2 conformal
tower.\cite{AffleckLudwig1} One finds, to $O(1/{\ell})$,
\begin{widetext}
\begin{equation}  \label{mChannelKondoSpectrum}
E = E_0 + \frac{\pi}{{\ell}} \left[ \left(\, \frac{1}{8} Q^2 + N_c\, \right) +
\left(\, \frac{j_s(j_s + 1)}{4}  + N_s\, \right) + \left(\, \frac{j_f(j_f+1)}{4}
+ 
N_f \, \right) \right] \, .
\end{equation}
\end{widetext}
Here $Q \in \Z$ are U(1) charge quantum numbers, while  $j_{s (f)} =0, 1/2, 1$
are quantum numbers for the level-2 SU(2) spin (flavor) primary states, 
with $N_c$, $N_s$, and $N_f$ labeling the 
corresponding descendant levels.
The quantum numbers are restricted to appear in the combinations
\begin{eqnarray}  \label{mChannelKondoGluing}
Q = 0 \ \mbox{mod 2},\  & j_s = \frac{1}{2},\  & j_f = 0 \ \mbox{or} \ 1, \nonumber
\\
Q = 1 \ \mbox{mod 2},\  & \ j_s = 0\ \ \mbox{or} \ 1,  \ & \ j_f = \frac{1}{2}, \, 
\end{eqnarray}
which define the {\em two-channel Kondo gluing conditions}
\cite{AffleckLudwig1} for the conformal towers. Since the charge quantum
numbers in Eqs. (\ref{mChannelKondoGluing}) are defined modulo 2, we can make a
shift $Q \rightarrow Q - 2$ in (\ref{mChannelKondoSpectrum}) without affecting
the gluing conditions. With this we recover the form of the conformal spectrum
as derived by Fujimoto and Kawakami \cite{FujimotoKawakami} from the exact
solution.\cite{AndreiDestri,TsvelikWiegmann} By mapping the nontrivial $Z_2$
part of the two-channel Kondo scattering onto that of a restricted
solid-on-solid model \cite{AndrewsBaxterForrester} coupled to the impurity,
these authors elegantly circumvented the ``string solution'' problem mentioned
above.

In exact analogy with the single-channel case detailed above, we can match the
finite-size two-channel Kondo spectrum defined by
by Eqs. (\ref{mChannelKondoSpectrum}) and (\ref{mChannelKondoGluing}) to that for the
$n_c \rightarrow 1$ limit of the two-channel Anderson model by performing a
shift $Q \rightarrow Q - 1$ in Eqs. (\ref{mChannelKondoSpectrum}) and
(\ref{mChannelKondoGluing}). The gluing conditions are accordingly modified
and now take the form
\begin{eqnarray}  \label{mChannelAndersonGluing}
Q = 0 \ \mbox{mod 2},\  &\ j_s = 0, \ \mbox{or} \ 1 \ & \ j_f = \frac{1}{2},
\nonumber \\
Q = 1 \ \mbox{mod 2},\  & j_s = \frac{1}{2},  & \ j_f = 0 \ \mbox{or} \ 1.  \, 
\end{eqnarray}
The gluing conditions (\ref{mChannelAndersonGluing}) can formally be obtained
by starting with the gluing conditions for free electrons (carrying spin and
flavor):
\begin{eqnarray} 
\label{FermiLiquidGluing}
Q = 0 \ \mbox{mod 2},\  &\ j_s = 0, \ & \ j_f = 0, \nonumber \\
                       & \ j_s = 1, \ & \ j_f = 1, \nonumber \\
Q = 1 \ \mbox{mod 2},  & j_s = \frac{1}{2},    & \ j_f = \frac{1}{2},  \, 
\end{eqnarray}
and then performing {\em Kac-Moody fusion with the $j_f = 1/2$ flavor
conformal tower:} $j_f = 0 \rightarrow 1/2; j_f = 1/2 \rightarrow 0, 1; j_f
= 1 \rightarrow 1/2$. Alternatively, we can obtain
Eqs. (\ref{mChannelAndersonGluing}) from Eqs. (\ref{FermiLiquidGluing}) by 
{\em fusion
with the $j_s = 1/2$ spin conformal tower: $j_s = 0 \rightarrow 1/2; j_s =
1/2 \rightarrow 0, 1; j_s = 1 \rightarrow 1/2$, concurrent with fusion with
the $Q=1$ charge conformal tower: $Q \rightarrow Q + 1$.}  In Sec.~(III.B)
we shall see how these two equivalent schemes can be given a simple physical
interpretation, considering the particular structure of the hybridization
interaction in (\ref{hybr}). It is important to realize that in contrast to
the single-channel case, the two-channel Kondo gluing conditions in
(\ref{mChannelKondoGluing}) cannot be traded for a local scattering potential
by the charge shift $Q \rightarrow Q - 1$. Although a local scattering
potential {\em is} generated [cf.~Eq.~(\ref{PotentialScattering}) below] the
``charge-renormalized'' gluing conditions (\ref{mChannelAndersonGluing}) are
{\em not} those of free electrons. A two-channel Kondo system forms a {\em
  non-Fermi liquid} with properties very different from those of phase-shifted
free electrons.

Let us now turn to the case of noninteger impurity valence $n_c \neq 1$.  As
supplementary input we use the fact that the low-lying excited states of the
model split into separate gapless charge, spin and flavor excitations for
any value of $ \varepsilon$ (and hence $n_c$), as revealed by the exact Bethe
Ansatz solution.\cite{BolechAndrei} Since there is no change of symmetry of
the model as we tune $\varepsilon_q$ and $\varepsilon_s$ away from the (magnetic
two-channel Kondo) limit $\varepsilon \ll \mu - \Gamma$, one
expects that a spectral U(1) charge, SU(2)$_2$ spin, and SU(2)$_2$ flavor
Kac-Moody structure is retained throughout the full range of magnetic and
quadrupolar energies. This implies that in the absence of any relevant
operators (that could push the model to a fixed point of a different Kac-Moody
structure), the finite-size spectrum must take the form
\begin{widetext}
\begin{equation}  \label{mChannelAndersonSpectrum}
E = E_0 + \frac{\pi}{{\ell}} \left[ \left( \frac{1}{8} ( Q - n_c)^2 -
\frac{1}{8}
n_c^2 + N_c \right) + \left( \frac{j_s(j_s + 1)}{4} + N_s \right) + 
\left( \frac{j_f(j_f+1)}{4} + N_f \right)
\right]
\end{equation}
\end{widetext}
with the values of the charge- spin- and flavor- quantum numbers constrained
by Eqs. (\ref{mChannelAndersonGluing}). Note that the charge levels have again been
shifted by a constant so that $Q$ measures the number of conduction electrons
added to or removed from the groundstate [i.e. $E_0 = E(Q=0)$].

Similar to the ordinary Anderson model, the charge spectrum in
(\ref{mChannelAndersonSpectrum}) implies the presence of an {\em effective}
local scattering potential
\begin{equation} \label{PotentialScattering}
V_{eff} = -\frac{n_c}{4} J(0) \, .
\end{equation}
The charge current $J$ has scaling dimension $\Delta = 1$, making $V_{eff}$ in
Eq. (\ref{PotentialScattering}) into an exactly marginal boundary operator with
the scaling field sampling the impurity charge valence $n_c$. Provided that no
relevant operators intervene under renormalization, it produces a line of
stable fixed points parameterized by the value of $n_c$ or, equivalently,
$\varepsilon$. This is in accordance with the Bethe
Ansatz solution,\cite{BolechAndrei} which established a line of
low-temperature fixed points analytically connected to the overscreened
two-channel (magnetic) Kondo fixed point at $ \varepsilon \ll \mu
- \Gamma$.

Given the finite-size spectrum defined by Eqs. (\ref{mChannelAndersonGluing}) and
(\ref{mChannelAndersonSpectrum}) we shall next identify the leading operator
content of the scaling theory, following the ``double-fusion'' prescription
outlined in Sec.~II. In particular, we must check that no relevant operators
are generated as we vary $ \varepsilon$ and move away from the integer
valence limit $n_c =1$.

Performing a second Kac-Moody fusion \cite{AffleckReview} with the $j_f = 1/2$
flavor conformal tower, the gluing conditions (\ref{mChannelAndersonGluing})
change to
\begin{eqnarray}  \label{DoubleFusionGluing}
Q = 0 \ \mbox{mod 2},\  & \ j_s = 0 \ \mbox{or} \ 1, \ & \ j_f = 0 \ \mbox{or} \
1, \nonumber \\
Q = 1 \ \mbox{mod 2},  & j_s = \frac{1}{2}, \  & j_f = \frac{1}{2}\, .   
\end{eqnarray}
These new gluing conditions ensure that the boundary condition 
in the half-plane geometry is time independent (corresponding to having 
attached an Anderson impurity to each edge,
$x=0$ {\em and} $x= \ell$, of the strip; cf. Sec. II). As we are now effectively
considering a finite-size spectrum with {\em two} quantum impurities present,
the charge quantum number gets renormalized twice, with $Q \rightarrow Q -
n_c(x=0)
\equiv Q^{\prime}$ from the impurity at the $x=0$ edge, and
$Q^{\prime} \rightarrow Q^{\prime} - n_c(x=\ell) = Q^{\prime} + n_c(x=0) = Q$
from the $x= \ell$ edge. Thus, the modifications of $Q$ caused by $n_c$ at each
edge
of the strip cancel each other, and the double-fusion spectrum becomes
independent of the impurity valence. The reason for the cancellation is the same
as in the
single-channel case, where the sum rule (\ref{FriedelLangreth}) connects $n_c$ 
to the scattering phase shift caused by the impurity. Since the phase shifts at
$x=0$
and $x= \ell$ have opposite signs (at one edge a left-moving electron is
reflected, while at the other edge a right-moving electron
gets reflected), one must formally assign charge valences of opposite signs
to the auxiliary impurities attached to $x=0$ and $x= \ell$, respectively. We
clearly
expect the same situation to apply also in the two-channel case.  
By comparison with Eqs. (\ref{CardyFormula}) and
(\ref{mChannelKondoSpectrum}) we thus read off for the possible boundary scaling
dimensions $\Delta = \Delta_c + \Delta_s + \Delta_f$:
\begin{gather}    
\Delta_c  =  \frac{1}{8} Q^2 + N_c \, , \nonumber \\
\Delta_s  =  \frac{1}{4} j_s(j_s+1) + N_s \, ,  \quad \quad
\Delta_f  =   \frac{1}{4} j_f(j_f+1) +N_f \label{ScalingDimensionSpectrum}
\end{gather}
with $Q$, $j_s$, and $j_f$ constrained by (\ref{DoubleFusionGluing}), and with
$N_c, N_s, N_f$ positive integers. 

The gluing conditions in Eqs. (\ref{DoubleFusionGluing}) are identical to those for
the spectrum of boundary scaling dimensions in the two-channel Kondo model.
\cite{AffleckLudwig1} Here we obtained them by double fusion with the $j_f =
1/2$ flavor conformal tower, while in the Kondo case they follow from double
fusion with the spin-1/2 conformal tower.  The equivalence of the two
procedures is consistent with the $Q \, \mbox{mod}\, 2$ invariance of
Eqs. (\ref{DoubleFusionGluing}). Double fusion with $j_f = 1/2$ in the empty
valence limit turns into double fusion with $j_s = 1/2$ in the Kondo limit
with a concurrent charge renormalization $Q \rightarrow Q + 2$. We shall
elaborate on this point in the next section.

The symmetries of the Hamiltonian in Eq. (\ref{hamiltonian}) impose
further restrictions on the allowed Kac-Moody charge, spin and
flavor quantum numbers. Charge conservation implies that $Q=0$ in
Eq. (\ref{ScalingDimensionSpectrum}) and hence only the trivial identity
conformal tower in the charge sector contributes to the spectrum of
boundary scaling dimensions. Moreover, since the charge valence $n_c$
does not appear in Eqs. (\ref{ScalingDimensionSpectrum}), no scaling
dimension can depend on it.  No relevant operator can therefore appear
as $n_c$ is tuned away from the magnetic two-channel Kondo limit. This
guarantees that the finite-size spectrum defined by Eqs. 
(\ref{DoubleFusionGluing}) and (\ref{ScalingDimensionSpectrum})
remains valid throughout the full range of charge valence $n_c \in (0,
1)$, with a line of stable fixed points parametrized by $n_c$. Recall
that this line is produced by the marginal operator $V_{eff}$ in
Eq. (\ref{PotentialScattering}), associated with the ``single-fusion''
spectrum, Eqs. (\ref{mChannelAndersonGluing}) and
(\ref{mChannelAndersonSpectrum}), which carries an explicit dependence
on $n_c$. $V_{eff}$ here plays the role of a {\em boundary changing}
operator.\cite{AffleckLudwig2} For a given value of $n_c$ it
determines the locus on the fixed line to which the theory flows under
renormalization. It may be tempting to conclude that the result
obtained from double fusion, Eqs. (\ref{DoubleFusionGluing}) and
(\ref{ScalingDimensionSpectrum}), implies that the time-independent
critical behavior is the {\em same} along the critical line, being
insensitive to $n_c$. This is correct with regard to {\em critical
exponents}, but as we shall see below, {\em amplitudes} of various
scaling operators pick up a dependence on $n_c$, giving rise to very
different physics as we move along the fixed line.

The leading boundary operator contributed by the charge sector is the exactly
marginal charge current $J(0) = \no{\psidop{\alpha \sigma}(0) \psiop{\alpha
\sigma}(0)}$. It is formally identified as the first Kac-Moody descendant
\cite{ItzyksonDrouffe} $J(0) = J_{-1} \1^{(c)}(0)$ of the identity operator
$\1^{(c)}$ (which generates the $Q=0$ conformal tower). As we have seen,
$J(0)$ being marginal, it shows up manifestly in the low-energy
spectral-generating Hamiltonian $H_{charge}$ in Eq. (\ref{SugawaraCharge}), and is
responsible for producing the line of fixed points.  The appearance of the
charge current breaks particle-hole symmetry, since $J(0) \rightarrow - J(0)$
under charge conjugation $\psiop{\alpha \sigma}(x) \rightarrow
\epsilon_{\alpha \beta} \epsilon_{\sigma \mu} \psidop{\beta \mu}(x)$. It is
interesting to note that while the same symmetry is broken by the ionic
pseudoparticle term (\ref{ion}) in (\ref{hamiltonian}) under $\fdop{\sigma}
\rightarrow \epsilon_{\sigma \mu} f_{\mu}$, the pieces (\ref{bulk}) and
(\ref{hybr}) of the Hamiltonian involving the conduction electron field
$\psiop{\alpha \sigma}(x)$ actually respect charge conjugation. This symmetry
gets broken dynamically via the coupling to the pseudo-particles, allowing the
charge current in Eq. (\ref{PotentialScattering}) to enter the stage.
\cite{FOOTNOTE}

The next-leading boundary operator from the charge sector is that of the
energy-momentum tensor, $T^{(c)}(0) = \mbox{$\no{J(0) J(0)}$}/4$, appearing as
the second {\em Virasoro descendant} \cite{ItzyksonDrouffe} of the identity
operator, $T_c(0) = L_{-2} \1^{(c)}(0)$. Although it carries scaling dimension
$\Delta = 2$ and hence is subleading to the charge current $J(0)$, being a
Virasoro descendant of the identity operator it has a nonvanishing
expectation value and turns out to produce the same scaling in temperature as
$J(0)$.  Although not central to the present problem, we shall briefly return
to this question in Sec.~IV.

Focusing now on the spin and flavor sectors, conservation of total spin and
flavor quantum numbers requires that all spin and flavor boundary operators
must transform as singlets. The leading spin boundary operator with this
property, ${\cal O}^{(s)}(0)$ call it, has dimension $\Delta_s = 3/2$ and is
obtained by contracting the spin-1 field ${\vphi}^{(s)}(0)$ (which generates the
$j_s =1$ conformal tower) with the vector of SU(2)$_2$ raising operators
$\vJ^{(s)}_{-1}: {\cal O}^{(s)}(0) = \vJ^{(s)}_{-1} \cdot {\vphi}^{(s)}(0)$. As
expected, this is the same operator that drives the critical behavior in the
two-channel (magnetic) Kondo problem.\cite{AffleckLudwig1} The next-leading
spin boundary operator is the energy momentum tensor $T^{(s)}(0) =
\no{\vJ^{(s)}(0) \cdot \vJ^{(s)}(0)}/4 \ [= L_{-2} \1^{(s)}(0)]$, of
dimension $\Delta_s = 2$.

The two leading flavor boundary operators are obtained in exact analogy with
the spin case. In obvious notation: ${\cal O}^{(f)}(0) = \vJ^{(f)}_{-1} \cdot
{\vphi}^{(f)}(0)$ of dimension $\Delta_f = 3/2$, and $T^{(f)}(0) =
\no{\vJ^{(f)}(0) \cdot \vJ^{(f)}(0)}\, [= L_{-2} \1^{(f)}(0) ]$ of dimension
$\Delta_f = 2$ .

Boundary operators with integer scaling dimensions generate an
analytic temperature dependence of the impurity thermodynamics,
\cite{AffleckReview} subleading to that coming from the two leading
irrelevant operators ${\cal O}^{(s)}(0)$ and ${\cal
O}^{(f)}(0)$. Hence, in what follows we shall focus on these latter
operators. In the case of the
two-channel Kondo problem the flavor operator ${\cal O}^{(f)}(0)$ was
argued to be effectively suppressed, at least for the case when the (bare)
Kondo coupling is sufficiently small.\cite{AffleckLudwig1} On
dimensional grounds one expects that the spin scaling field
$\lambda_s$ [multiplying ${\cal O}^{(s)}(0)$ in the scaling
Hamiltonian] is ${\mathit O}(1/\sqrt{T_K})$, where $T_K$ is the Kondo
scale for the crossover from weak coupling (high-temperature phase) to
strong renormalized coupling (low-temperature phase). The flavor
operator, on the other hand, does not see this scale since the
infrared divergences in perturbation theory (which signal the
appearance of a dynamically generated scale) occur only in the spin
sector. The only remaining scale is that of the band width $D$ (which
plays the role of an ultraviolet cutoff), implying that the flavor
scaling field $\lambda_f$ [multiplying ${\cal O}^{(f)}(0)]$ is
${\mathit O}(1/\sqrt{D})$.  For a small (bare) Kondo coupling $\lambda$, $T_K
\sim D\, \mbox{exp} (-1/\lambda) \ll D$, and the critical behavior is
effectively determined by ${\cal O}^{(s)}(0)$ alone.

As we will show, the picture changes dramatically for the two-channel
Anderson model. The Bethe Ansatz solution\cite{BolechAndrei} shows that there
are {\em two} dynamically generated temperature scales $T_{\pm}(\varepsilon)$
present in this model. They determine and parametrize the 
thermodynamic response of the system.  Consider for example the quenching of
the entropy as the temperature is lowered: in the magnetic moment regime
$(\varepsilon \ll \mu - \Gamma)$ the two scales are widely separated, $T_{+}
\ll T_{-}$.  For temperatures in the range $T_+ \le T \le T_-$ charge and
flavor fluctuations are suppressed and the entropy shows a plateau
corresponding to the formation of a local moment. The moment undergoes
frustrated screening when the temperature is lowered below $T_-$ reaching zero
point entropy $S=k_B \ln \sqrt{2}$, and $n_c =1$. As $\varepsilon$ decreases
$T_{+}$ and $T_{-}$ approach each other and the range over which the magnetic
moment exists decreases too. It disappears (i.e.~$T_+=T_-$) when $\varepsilon =
\mu$. At this point no magnetic moment forms. However the system still flows
to a frustrated two-channel Kondo fixed point, with entropy $S=k_B \ln \sqrt{2}$,
but with $n_c=1/2$.  Continuing along the critical line, the two scales trade
places, and eventually, at the quadrupolar critical endpoint $(\varepsilon \gg
\mu - \Gamma)$, one finds that $T_{+} \gg T_{-}$.  Since the irrelevant
spin (flavor) operator is expected to dominate the critical behavior at the
magnetic (quadrupolar) fixed point, we are led to conjecture that the
corresponding scaling field is parametrized precisely by $1/\sqrt{T_{+}}$
$(1/\sqrt{T_{-}})$. With this scenario played out, the relative importance of
the two
boundary operators changes continuously as one moves along the fixed line,
with each operator ruling at its respective critical end point.

In the next section we shall prove our conjecture by computing the
impurity specific heat and susceptibility following from the scaling
Hamiltonian and comparing it to the low temperature thermodynamics from the
exact Bethe Ansatz solution.\cite{BolechAndrei} We shall also use this 
comparison to ``fit'' the
parametrization of the scaling fields. This fixes the critical theory
completely.

\subsection{Application: Fermi Edge Singularities}

Before taking on this task, however, we shall exploit the BCFT scheme
developed above to determine the scaling dimensions of the pseudoparticles
which enter the Hamiltonian in (\ref{ion}) and (\ref{hybr}). The analogous
problem for the ordinary Anderson model has been extensively studied
\cite{MengeMullerHartmann,Costi,FujimotoKawakamiYang3}, in part because of its
close connection to the {\em X-ray absorption problem}.

One is here interested in the situation where X-ray absorption knocks
an electron from a filled inner shell of an ion in a metal into the
conduction band. As an effect the conduction electrons experience a
transient local potential at the ion which lost the core electron, and
this typically produces a singularity in the X-ray spectrum $\sim
(\omega - \omega_F)^{-\beta}$ close to the Fermi level threshold
$\omega_F$.  Nozieres and de Dominicis,\cite{NozieresDominicis} using
a simple model, showed that the exponent $\beta$ depends only on the
phase shift which describes the scattering of conduction electrons
from the core hole created by the X-ray.  Fermi edge singularities
linked to local dynamic perturbations are fairly generic,
\cite{Toulouse} and it is therefore interesting to explore how they
may appear in a slightly more complex situation where a localized
``deep hole'' type perturbation involves additional degrees
of freedom. Indeed, the two-channel Anderson model in
Eq. (\ref{hamiltonian}) provides an ideal setting for this as it
accommodates a local quantum impurity carrying both spin {\em and}
channel (quadrupolar) degrees of freedom.

To set the stage, let us make a {\em Gedankenexperiment} and imagine that we
suddenly replace a thorium ion in the host metal (say,
U$_{0.9}$Th$_{0.1}$Be$_{13}$) 
by a uranium
ion, the sole effect being to introduce a localized quadrupolar {\em or}
magnetic moment. To describe the response of the host to this perturbation, we
consider the pseudoparticle propagators
\begin{eqnarray}
G_{\sigma} & = & \bra{0} e^{iHt} \fop{\sigma} e^{-iHt} \fdop{\sigma} \ket{0}
\label{f} \\
 G_{\bar{\alpha}} & = & \bra{0} e^{iHt} \bop{\bar{\alpha}} e^{-iHt} 
\bdop{\bar{\alpha}} \ket{0} 
  \label{b}
\end{eqnarray}
(with no summation over $\alpha, \sigma$). At time $t < 0$ the ground state
$\ket{0}$ of the metal is that of free conduction electrons. At $t=0$ a
pseudoparticle (creating the localized moment) is inserted into the metal,
and the system evolves in time governed by the two-channel Anderson
Hamiltonian in Eq. (\ref{hamiltonian}). At some later time $t$ the pseudo-particle
is removed, and one measures the overlap between the state obtained with the
initial unperturbed groundstate. It is important to realize that by inserting
(removing) a pseudoparticle operator, one simultaneously turns on (off) the
impurity terms in Eqs. (\ref{ion}) and (\ref{hybr}), and hence changes the dynamics
of the conduction electrons. As we reviewed in Sec.~(II), at low temperatures
and close to a critical point, this corresponds to a change of the boundary
condition which emulates the presence (absence) of the impurity. Hence, the
pseudoparticle operators $f_{\sigma}$ and $b_{\alpha}$ are {\em boundary
changing operators},\cite{Cardy} and we can use the BCFT machinery to
calculate their propagators in Eqs. (\ref{f}) and (\ref{b}). Our discussion closely
follows that in Ref.~\onlinecite{AffleckLudwig2}.

We start by considering the free-electron theory defined on the complex upper
half-plane $\Cplus = \{z = \tau + ix; x \ge 0 \}$ with a trivial (Fermi
liquid) boundary condition, call it $``A''$, imposed at the real axis $x =0$.
We denote by $\ket{A; 0}$ the ground state for this configuration. By injecting
a pseudoparticle at time $\tau =0$, the boundary condition for $\tau > 0$
changes to $``B''$, here labeling the boundary condition which corresponds to
the nontrivial gluing condition in (\ref{mChannelAndersonGluing}) for the
spectrum of the two-channel Anderson model.  By mapping the half-plane to a
strip $\{w = u + iv; 0 \le {\ell} \}$ via the conformal transformation $w =
({\ell}/2\pi) \ln z $, the boundary of the strip at $v = {\ell}$ also turns
into type $B$ after insertion of the pseudoparticle operator.  Under the same
transformation the pseudoparticle propagator in 
$\Cplus$, $\bra{A; 0} {\cal
O}_i(\tau_1) {\cal O}^{\dagger}_i(\tau_2) \ket{A;0} = (\tau_1 - \tau_2)^{-
2x_i}$ (with $x_i$ the dimension of ${\cal O}_i, i = f$ or $b$) transforms as
\begin{widetext}
\begin{equation}    \label{PseudoCorrelation}
(\tau_1 - \tau_2)^{-2x_i}  \rightarrow  
\left( \, \frac{2{\ell}}{\pi} \sinh \frac{\pi}{2{\ell}}(u_1 -
u_2) \, \right)^{-2x_i} = \sum_n \mid \bra{AA;0} {\cal O}_i(0) \ket{AB;n} 
\mid^2 e^{-(E_n^{AB} - E_0^{AA}) (u_2 -u_1)} \, ,
\end{equation}
\end{widetext}
where on the right-hand side of Eq. (\ref{PseudoCorrelation}) we have inserted a complete
set of states $\ket{AB;n}, n = 0,1,...$ of energies $E_n^{AB}$, defined on the
strip with boundary condition $A \ (B)$ at $u=0 \ (u={\ell})$.  In the limit
$u_2 - u_1 \gg {\ell}$:
\begin{equation}        \label{AsymptoticCorrelation}
\left(\, \frac{2{\ell}}{\pi} \sinh \frac{\pi}{2{\ell}}(u_1 - u_2)\,
\right)^{-2x_i} \rightarrow 
\frac{\pi}{{\ell}} e^{-[\pi x_i(u_2 -u_1)/\ell]} \, .
\end{equation}
It follows, by comparison with Eq. (\ref{PseudoCorrelation}), that
\begin{equation}       \label{PseudoDimension}
x_i = \frac{{\ell}}{\pi} (E_{n_i}^{AB} - E_0^{AA}) \, ,
\end{equation}
where $n_i$ is the lowest energy state with a nonzero matrix element in the
sum in Eq. (\ref{PseudoCorrelation}).

The pseudoparticle spectrum $I_i(\omega)$ close to the Fermi level is
obtained by Fourier transforming the corresponding propagator in Eq. (\ref{f}) or
(\ref{b}), and it follows that
\begin{equation} \label{Spectral}
I_i(\omega) \sim \frac{1}{\mid \omega - \omega_F \mid^{1-2x_i}} \, , \ \ \ \ \ \
\  i = f, b,
\end{equation}
with a singularity when $x_i < 1/2$. 

Let us consider first the slave boson $b_{\alpha}$, carrying quantum numbers
$Q=0, j_s=0, j_f = 1/2$. Its scaling dimension $x_b$ is given by
Eq. (\ref{PseudoDimension}) where $E_{n_i}^{AB}$ is obtained from
Eq. (\ref{mChannelAndersonSpectrum}) by putting $Q=0, j_s=0, j_f = 1/2$, and with
$E_0^{AA} = 0$ (with proper normalization of energies).  To be able to compare
energies corresponding to different boundary conditions it is important to
keep the overall energy normalization fixed, and hence we must remove the
normalization constant $-\pi n^2_c/8\ell$ in Eq. (\ref{mChannelAndersonSpectrum})
before reading off the answer.  (Recall that this normalization
constant is expressly designed to remove the $Q=0$ contribution from the
charge sector to the ground state energy.) We thus obtain
\begin{equation} \label{SlaveBoson}
x_b = \frac{3}{16} + \frac{1}{8}n_c^2 \, .
\end{equation}
Similarly, we obtain for the pseudo fermion $(Q=1, j_s = 1/2, j_f = 0)$:
\begin{equation} \label{PseudoFermion}
x_f = \frac{5}{16} - \frac{1}{4}n_c + \frac{1}{8}n_c^2 \, .
\end{equation}

It is interesting to compare Eqs. (\ref{SlaveBoson}) and
(\ref{PseudoFermion}) with the corresponding scaling dimensions in the
ordinary Anderson model, obtained in
Refs.~\onlinecite{MengeMullerHartmann,FujimotoKawakamiYang3}:
$x^{\prime}_b = n_c^2/2, x^{\prime}_f = 1/2 - n_c/2 + n_c^2/4$. We
conclude that by opening an additional channel for the electrons, the
singularity in the slave boson spectrum in Eq. (\ref{Spectral})
(corresponding to X-ray photoemission) gets softer, whereas for the
pseudofermion spectrum (corresponding to X-ray absorption) the
opposite is true. One can show that this trend is systematic, and gets
more pronounced as the number of channels increases.\cite{JAB} We also
note that the values of the pseudoparticle scaling dimensions as
obtained by various approximation schemes, e.g. the
''non-crossing approximation'' (NCA) and large-$N$
calculations, \cite{CoxRuckenstein} are not in agreement with the
exact results presented here.

Before closing this section, let us return to the {\em Gedankenexperiment}
where we inserted an empty (degenerate) impurity level by adding a slave boson
to the filled Fermi sea. The slave boson carries only flavor $j_f = 1/2$, and
it follows that the allowed values of the quantum numbers of the combined
system are obtained by coupling $j_f = 1/2$ to the flavor quantum numbers
$j_f^{\prime}$ of the conduction electrons (leaving charge and spin
untouched).  Since $j_f$ and $j_f^{\prime}$ label conformal towers, this
coupling is {\em precisely governed by conformal fusion with} $j_f = 1/2:
j_f^{\prime} \rightarrow \, \mid j_f^{\prime} - 1/2 \!\mid, \mid
\!j_f^{\prime} -
1/2 \!\mid +1,... \,\mbox{min} (j_f^{\prime} + 1/2, 3/2 - j_f^{\prime})$.  By
letting the system relax, a non-zero charge valence $n_c$ may then build up at
the impurity site (as determined by the particular value of $\varepsilon =
\varepsilon_s - \varepsilon_q$). Formally, this process is captured by the
charge
renormalization $Q \rightarrow Q - n_c$ in the finite size-spectrum
(\ref{mChannelAndersonSpectrum}), taking place concurrently with the fusion in
flavor space which produces the nontrivial gluing conditions
(\ref{mChannelAndersonGluing}). Running the {\em Gedankenexperiment} with a
pseudo-fermion instead (carrying charge and spin, but no flavor) leads to a
scenario dual to the one above, where the coupling is now governed by fusion
with $j_s = 1/2$ in the spin sector and $Q=1$ in the charge sector [cf.~the
discussion after Eq.~(\ref{FermiLiquidGluing})].  As the system relaxes an
average charge $n_c$ goes to the impurity site, again leading to the
renormalization $Q \rightarrow Q - n_c$ in the spectrum
(\ref{mChannelAndersonSpectrum}). We conclude that our key results, Eqs.
(\ref{mChannelAndersonGluing}) and (\ref{mChannelAndersonSpectrum}) --
obtained in Sec (II.A) via a formal analysis -- can be given a natural
interpretation by considering the pseudoparticle structure of the
hybridization interaction in Eq. (\ref{hybr}).

\section{Low-Temperature Thermodynamics}

\subsection{Zero-Temperature Entropy}

The Bethe Ansatz solution \cite{BolechAndrei} of the model reveals that the
line of fixed points is characterized by a zero-temperature entropy $S_{imp} =
k_B \ln\sqrt{2}$ typical of the two-channel Kondo fixed point. As we shall show
next, this property emerges naturally when feeding in our result from
Sec.~(II.A) into the general BCFT formalism developed by Cardy.\cite{Cardy}

Recall that the key ingredient in this formalism is the {\em modular
  invariance} \cite{DiFrancesco} of a conformal theory. Applied to a model
defined on a cylinder of circumference $\beta = 1/T$ and length $\ell$, and
with conformally invariant boundary conditions $A$ and $B$ imposed at the open
ends, this means that its partition function $Z_{AB}$ is invariant under
exchange of space and (Euclidean) time variables. In our case, $B$ is the
boundary condition obtained from $A$ by attaching an Anderson impurity to one
of the edges of the cylinder (with $A$ a trivial Fermi liquid boundary
condition, to be defined below).  Now, let $H^{\ast}$ be the critical
Hamiltonian that corresponds to the spectrum given by
Eqs. (\ref{mChannelAndersonGluing}) and (\ref{mChannelAndersonSpectrum}). Then
$H_{AB} \equiv (\ell/\pi)H^{\ast}$ generates time translations around the
cylinder specified above. The partition function $Z_{AB}$ can hence be written
\begin{equation}     \label{Partition1}
Z_{AB} = \mbox{Tr e}^{-\beta H_{AB}} = \sum_a n^a_{AB} \chi_a(e^{-\pi
\beta/\ell}) \, .
\end{equation}
Here $\chi_a$ is a character of the conformal U(1) $\times$ SU(2)$_2 \times$
SU(2)$_2$ algebra, with $a = (Q, j_s, j_f)$ labeling a product of charge,
spin, and flavor conformal towers [constrained by
Eqs. (\ref{FermiLiquidGluing})], and with $n^a_{AB}$ counting its degeneracy
within the spectrum of $H_{AB}$. By a modular transformation $\beta
\leftrightarrow \ell$ the role of space and time is exchanged, and the
partition function now gets expressed as
\begin{equation}        \label{Partition2}
Z_{AB} = \bra{A} e^{- \ell H_{\beta}} \ket{B} \, ,
\end{equation}
where $H_{\beta}$ is the transformed Hamiltonian generating translations {\em
along} the cylinder, in the space direction, with $\ket{A}$ and $\ket{B}$
{\em boundary states} \cite{Cardy} implicitly defined by Eq. (\ref{Partition2}).
By inserting a complete set of {\em Ishibashi states} \, $\ket{a} = \sum_m
\ket{a;m}_L \otimes \ket{a;m}_R$ into Eq. (\ref{Partition2}) (with $m$ running
over all states in an $L/R$ product of charge, spin, and flavor conformal
towers labelled by the index $a$, one obtains \cite{DiFrancesco}
\begin{equation}    \label{Partition3}
Z_{AB} = \sum_a \cf{A\! \mid\! a} \cf{a\! \mid\! B} \chi_a (e^{-4\pi
\ell/\beta}) \, .
\end{equation}
Equating Eqs. (\ref{Partition1}) and (\ref{Partition3}) one can immediately read
off {\em Cardy's equation} \cite{Cardy}
\begin{equation}     \label{CardyEquation}
\sum_b n^b_{AB} S_{ab} = \cf{A\! \mid\! a} \cf{a\! \mid\! B} \, ,
\end{equation}
where $S_{ab}$ is the {\em modular S matrix} defined by
\begin{equation} \label{Modular}
\chi_a(e^{-\pi \beta/\ell}) = \sum_b  S_{ab} \chi_b(e^{-4\pi \ell/\beta}) \, ,
\end{equation}
with $a$ and $b$ indexing products of charge, spin, and flavor conformal
towers. It follows that the nontrivial boundary state $\ket{B}$ (which
emulates the presence of the impurity) is related to $\ket{A}$ via the
identity \cite{Cardy}
\begin{equation}   \label{FusionConnection}
\cf{a\! \mid\! B} = \cf{a\! \mid\! A} \frac{S_{ad}}{S_{a0}} \, ,
\end{equation}
with 0 labeling the product of identity conformal towers and $d$ the product
of conformal towers with which the Kac-Moody fusion of the finite-size
spectrum is performed.  As we found in Sec.~(II.A), for the two-channel
Anderson model, $d = (Q\! =\! 0, j_s \!=\!  0, j_f \!=\! 1/2)$. Factoring the
modular S matrices in charge, spin, and flavor, with $a = (Q, j_s, j_f)$
constrained by Eqs. (\ref{FermiLiquidGluing}) but otherwise arbitrary, one obtains
\begin{equation}   \label{ModularFactors}
\frac{S_{ad}}{S_{a0}} = \frac{S^c_{Q 0} S^s_{j_s 0} S^f_{j_f 1/2}}{S^c_{Q 0}
S^s_{j_s 0}
S^f_{j_f 0}} = \frac{S^f_{j_f 1/2}}{S^f_{j_f 0}} \, .
\end{equation}
The modular S matrix for the SU(2)$_2$ flavor symmetry is given by
\cite{DiFrancesco}
\begin{equation}   \label{FlavorModular}
S_{j_f j^{\prime}_f} = \frac{1}{\sqrt{2}} \sin 
\left[ \frac{\pi (2j_f + 1)(2 j^{\prime}_f + 1)}{4}\right]\, ,
\end{equation}
and it follows from Eqs. (\ref{FusionConnection}) and (\ref{ModularFactors}) that
\begin{equation}   \label{FlavorFusionConnection}
\cf{a\! \mid\! B} = \cf{a\! \mid\! A} \frac{\sin[\frac{\pi}{2} (2j_f + 1)]}
{\sin[\frac{\pi}{4} (2j_f + 1)]} \, .
\end{equation}
We now have the tools for deriving the impurity entropy. Following the
analogous analysis of the multichannel Kondo model in Ref.~\onlinecite{AffleckLudwig3}
we take the limit $\ell/\beta \rightarrow \infty$ in Eq. (\ref{Partition3}) (i.e.
we take $\beta \rightarrow \infty$ {\em after} the large-volume limit $\ell
\rightarrow \infty$). As a result, only the character of the groundstate,
$\chi_{a = (0,0,0)}(e^{-4\pi \ell/\beta}) = e^{\pi \ell c /6 \beta}$,
contributes to $Z_{AB}$ (with $c$ the conformal anomaly of the theory). One
thus obtains
\begin{equation}    \label{Groundstate}
Z_{AB} \rightarrow  e^{\pi \ell c /6 \beta} \cf{A\! \mid\! 0} \cf{0\! \mid\!
  B} \, , \ \ \ \ \ \  \ell/\beta
\rightarrow \infty ,
\end{equation}
with $\ket{0} \equiv \,\, \ket{Q=0, j_s =0, j_f =0}$. The free energy is thus
\begin{equation}      \label{FreeEnergy}
F_{AB} = -\pi c k_BT^2 \ell/6 - k_BT \ln (\cf{A \mid 0} \cf{0 \mid B}).   
\end{equation}
The second term in Eq. (\ref{FreeEnergy}) is independent of the size $\ell$ of the
system, and we therefore identify
\begin{equation}     \label{ImpurityEntropy1}
S_{imp} = k_B \ln (\cf{A\! \mid\! 0} \cf{0\! \mid\! B})
\end{equation}
as the impurity contribution to the zero-temperature entropy. With no impurity
present, $A=B$ and $S_{imp} = 0$, and it follows that $\ket{A} = \,\, \ket{0}$
[consistent with the {\em Fermi liquid gluing condition} in
Eq. (\ref{FermiLiquidGluing})]. Thus, from Eqs. (\ref{FlavorFusionConnection}) and
(\ref{ImpurityEntropy1}) we finally get
\begin{equation}      \label{ImpurityEntropy2}
S_{imp} = k_B \ln \sqrt{2}  \, ,
 \end{equation}
in agreement with the Bethe Ansatz result in Ref.~\onlinecite{BolechAndrei}.

A few remarks may here be appropriate. First, we note that the effective
renormalization of the charge sector, $Q \rightarrow Q - n_c$
[cf.~Eq.~(\ref{mChannelAndersonSpectrum})], does {\em not} influence the
impurity entropy. At first glance, this may seem obvious: the degeneracies of
the impurity levels $-$ and hence the impurity entropy $-$ is {\em a priori} not
expected to be affected by the potential scattering in
(\ref{PotentialScattering}) caused by the impurity valence $n_c$.  The picture
becomes less trivial when one realizes that the impurity valence in fact
controls {\em how} the spin and flavor degrees of freedom get screened by the
electrons.  As found in the Bethe Ansatz solution \cite{BolechAndrei}, for
nonzero $\varepsilon = \varepsilon_s - \varepsilon_q$, the quenching of the
entropy
occurs in two stages as the temperature is lowered.  For sufficiently large
values of $\mid \varepsilon - \mu \mid$, the free impurity entropy $k_B \ln 4$
is
first reduced to $k_B \ln 2$ at an intermediate temperature scale, suggestive
of a single remnant magnetic {\em or} quadrupolar moment (depending on the
sign of $\varepsilon$ or, equivalently, the sign of $n_c - 1/2$).  As the
temperature is lowered further, this moment gets overscreened by the
conduction electrons, as signaled by the residual entropy $k_B \ln \sqrt{2}$.
Remarkably, as found in Ref.~\onlinecite{BolechAndrei}
and verified here within the BCFT
formalism, the {\em same} residual impurity entropy $k_B \ln \sqrt{2}$
appears also in the mixed-valence regime $\mid \varepsilon - \mu \mid \, \le
\Gamma$ (where $n_c \approx 1/2$), for which there is no single localized
moment present at any temperature scale.

\subsection{Impurity Specific Heat}

We now turn to the calculation of the impurity specific heat. Again, the
analysis closely parallels that for the two-channel Kondo model.
\cite{AffleckLudwig1} Some elements are new, however, and we here try to
provide a self-contained treatment.

First, we need to write down the scaling Hamiltonian $H_{scaling}$ that
governs the critical behavior close to the line of boundary fixed points. To
leading order it is obtained by adding the dominant boundary operators found
in Sec (III) to the bulk critical Hamiltonian $H^*$ [representing $H_{bulk}$
in Eq. (\ref{hamiltonian})]:
\begin{multline}    \label{ScalingHamiltonian}
H_{scaling} = H^* + \lambda_c J(0) + \lambda_s {\cal O}^{(s)}(0)
+ \lambda_f {\cal O}^{(f)}(0) \\
+ \mbox{subleading terms} \, ,
\end{multline}
with $\lambda_{c,s,f}$ the corresponding conjugate scaling fields. As
reflected in the finite-size spectrum (\ref{mChannelAndersonSpectrum}), $H^*$
splinters into dynamically independent charge, spin, and flavor pieces,
implying, in particular, that all correlation functions decompose into
products of independent charge, spin and flavor factors. {\em On} the
critical line the impurity terms $H_{ion}$ and $H_{hybr}$ in Eqs.~(\ref{ion}) and
(\ref{hybr}) together masquerade as a boundary condition on $H^*$, coded in
the nontrivial gluing conditions in Eqs.~(\ref{mChannelAndersonGluing}).  In
addition $H_{ion}$ and $H_{hybr}$ also give rise to the exactly marginal
charge current term $\lambda_c J(0)$ in Eq.~(\ref{ScalingHamiltonian}).  This term
keeps track of the charge valence acquired by the impurity.  As we have
already noticed, this is different from the Kondo problem where the local spin
exchange interaction at criticality gets {\em completely} disguised as a
conformal boundary condition.\cite{AffleckReview} The reason for the
difference is that away from the (integer valence) Kondo limit, the effective
local scattering potential in Eq.~(\ref{PotentialScattering}) cannot be removed by
redefining the charge quantum numbers $Q \rightarrow Q-2$ in
Eqs.~(\ref{mChannelAndersonGluing}) and (\ref{mChannelAndersonSpectrum}). This in
turn reflects the fact that for non-integer valence the dynamics manifestly
breaks particle-hole symmetry, whereas for integer valence there is an
equivalent description \cite{BolechAndrei} in terms of the two-channel Kondo
model where this symmetry is restored \cite{FOOTNOTE}.  {\em Off} the critical
line the impurity terms in Eqs.~(\ref{ion}) and (\ref{hybr}) appear in the guise of
irrelevant boundary terms, of which $\lambda_s {\cal O}^{(s)}(0)$ and
$\lambda_f {\cal O}^{(f)}(0)$ in Eq.~(\ref{ScalingHamiltonian}) are the leading
ones.

Recall from Sec.~(II) that a boundary critical theory by default is defined in
the complex half-plane $\Cplus = \{\im z > 0 \}$, with the boundary condition
imposed on the real axis $\im z=0$.  As we have seen, by continuing the fields
to the lower half-plane we obtain a {\em chiral} representation of this
theory, now defined in the full complex plane $\C$. Via the transformation $w=
\tau + ix = (\beta/\pi) \, \mbox{arctan}(z)$, $\C$ can be mapped conformally
onto an infinite cylinder, $\Gamma_c$ call it, of circumference $\beta$:
$\Gamma_c
= \{ (\tau, x) ; -\beta/2 \leqslant \tau \leqslant \beta/2; -\infty \leqslant
x \leqslant \infty \}$.  It is convenient to use $\Gamma_c$ as a
``finite-temperature geometry'' by treating $\beta$ as an inverse temperature.
The partition function can then be written as
\begin{widetext}
\begin{equation}   \label{Partitionfunction}
e^{-\beta F(\beta,\{\lambda_{\nu}\})} =  e^{-\beta F(\beta,0)} \,
\cf{\, \mbox{exp}\, ( \int_{-\beta/2}^{\beta/2} d\tau [ \lambda_c {\tilde
 J}(\tau, 0)  
+ \lambda_s {\tilde
{\cal O}}^{(s)}(\tau, 0) + \lambda_f {\tilde {\cal O}}^{(f)}(\tau, 0)\, ]  
+ \mbox{subleading terms}\, ) \,}_T , 
\end{equation}
where $ F(\beta,\{\lambda_{\nu}\}) = F_{bulk}(\beta, 0) +
f_{imp}(\beta,\{\lambda_{\nu}\}) $ is the sum of bulk and impurity free
energies, with $\{\lambda_{\nu}\} \equiv (\lambda_c, \lambda_s,
\lambda_f,...)$ the collection of boundary scaling fields. We have here passed
to a Lagrangian formalism, with the ``tilde'' and $\cf{\ }_T$ referring to the
finite-$T$ geometry $\Gamma_c$. By a linked cluster expansion we obtain from
Eq.~(\ref{Partitionfunction}) the impurity contribution to the free energy:
\begin{equation}   \label{LinkedCluster}
f_{imp}(\beta, \{\lambda_{\nu}\}) =  -\frac{1}{2\beta}
\int \! \! \int_{-\beta/2}^{\beta/2} d\tau_1 \, d\tau_2
[\, \lambda_c^2 \cf{{\tilde J}(\tau_1, 0) {\tilde J}(\tau_2, 0)}_{T, c} +
\sum_{\nu=s,f}
 \lambda_{\nu}^2 \cf{{\tilde {\cal O}}^{\nu}(\tau_1, 0) 
{\tilde {\cal O}}^{\nu}(\tau_2, 0)}_{T, c}\, ] +  \mbox{subleading terms} \, , 
\end{equation}
\end{widetext}
with $\cf{\ }_{T,c}$ denoting a connected two-point function in $\Gamma_c$ and
where, for a fixed temperature $1/\beta$, the constant term $f_{imp}(\beta,
0)$ has been subtracted. Note that the linear terms in the expansion are
absent since all three boundary operators are Virasoro primary
\cite{DiFrancesco} and hence have vanishing expectation values. Since
two-point functions containing operators from different sectors (charge, spin
or flavor) decompose into products of one-point functions, these also vanish.
Similarly, connected two-point functions reduce to ordinary two-point
functions.  Also note that there is no term in Eq.~(\ref{LinkedCluster}) of zeroth
order in the scaling fields.  This is a generic feature of quantum impurity
problems, which is best understood by looking at the finite-size scaling form
\cite{CardyBook} for the impurity free energy:
\begin{equation}     \label{ScalingAnsatz}
f_{imp}(\beta, \{\lambda_{\nu}\}) = E_{imp} + \frac{1}{\beta}Y_{imp}
(\{\lambda_{\nu}\beta^{y_{\nu}} \}) + \dots \, ,
\end{equation}
with $Y_{imp}$ a universal scaling function and $E_{imp}$ nonuniversal.  The
exponents $y_{\nu}$ are RG eigenvalues, connected to the dimensions
$\Delta_{\nu}$ of the irrelevant boundary operators by $y_{\nu}
=1-\Delta_{\nu}$. It follows immediately from Eq.~(\ref{ScalingAnsatz}) that {\em
at} the fixed point
\begin{equation}  \label{CimpCrit}
C_{imp}(\{\lambda_{\nu} = 0 \}) = -T \frac{\partial^2 f_{imp}}{\partial T^2} =
0.
\end{equation}
Thus, to see the effect of the impurity, the leading irrelevant boundary
operators have to be added explicitly, and these operators then produce the
dominant terms in the scaling in temperature.

The correlation function of chiral Virasoro primary operators ${\cal O}^{(i)}$
in $\C$ takes the familiar form
\begin{equation}   \label{TwoPoint}
\cf{{\cal O}^{(i)}(z_1) {\cal O}^{(i)}(z_2)} =
\frac{A_i}{(z_1-z_2)^{2\Delta_i}}.
\end{equation}
Here $\Delta_i$ is the scaling dimension of ${\cal O}^{(i)}$, and $A_i$ is a
normalization constant. Mapping the real axis $\im z=0$ into $\Gamma_c$, this
transforms into
\begin{equation}  \label{TwoPointTemperature}
\cf{\tilde{{\cal O}}^{(i)}(\tau_1,0)\tilde{{\cal O}}^{(i)}(\tau_2,0)}_T \ = \
\frac{A_i}{ |\frac{\beta}{\pi}\mbox{sin}
(\frac{\pi}{\beta}(\tau_1 - \tau_2))|^{2\Delta_i}} \ {} \, ,
\end{equation}
and it follows that Eq.~(\ref{LinkedCluster}) can be rewritten as
\begin{widetext}
\begin{equation}  \label{ImpurityIntegral}
f_{imp}(\beta, \{\lambda_{\nu}\})  =  -\frac{1}{2\beta}
\int_{\mbox{tan} \kappa}^{\infty} du \,
\left(\, \lambda_c^2 A_c \frac{1}{u^2} + ( \lambda_s^2 A_s + \lambda_f^2 A_f )
\frac{\sqrt{1+u^2}}{u^3} \, \right) + \mbox{subleading terms} \, ,
\end{equation}
\end{widetext}
with $\kappa$ a UV cutoff. This form is obtained by replacing the double
integral in Eq.~(\ref{LinkedCluster}) by a single integral over $\tau = \tau_1 -
\tau_2$, exploiting that the two-point functions in
Eq.~(\ref{TwoPointTemperature}) are periodic in $\tau$ (with period $\beta$).
Introducing the auxiliary variable $u \equiv \mbox{tan} (\pi \tau / \beta)$
and specifying the scaling dimensions $\Delta_c = 1, \Delta_s = \Delta_f =
3/2$ in Eq.~(\ref{TwoPointTemperature}), one arrives at the expression in
Eq.~(\ref{ImpurityIntegral}) after inserting the short-time cutoff $\tau_0 =
\kappa \beta /\pi$. The integral can be evaluated straightforwardly
\cite{AffleckLudwig1,FrojdhJohannesson} and in the limit of low temperatures,
i.e. for small $\kappa$, an expansion of ${\mbox{tan}(\kappa})$ yields for the
impurity specific heat:
\begin{widetext}
\begin{multline}  \label{ImpuritySpecificHeat}
C_{imp}  =  -T \frac{\del^2 f_{imp}}{\del T^2}  
         =  (\lambda_s^2 A_s + \lambda_f^2 A_f)\pi^2 T \ln
(\frac{1}{\tau_0 T}) + \lambda_c^2 A_c \pi^2 \tau_0 T \\ 
         +  \mbox{corrections from subleading boundary operators} \, .
\end{multline}
\end{widetext}
Before proceeding, let us comment on the correction terms in
Eq.~(\ref{ImpuritySpecificHeat}).  The next-leading boundary operators are the
energy-momentum tensors $T^{(c,s,f)}$ identified in Sec.~(III.A). Passing to the
finite-$T$ geometry $\Gamma_c$, their contribution to the partition function
takes the form
\begin{displaymath}
e^{-\beta F(\beta,0)} \, 
\cf{\, \mbox{exp}\, ( \sum_{i=c,s,f}  \int_{-\beta/2}^{\beta/2} d\tau 
\lambda^{\prime}_i {\tilde T}^{(i)}(\tau, 0) ) \,}_T \, ,
\end{displaymath}
with $\lambda_i^{\prime}$ conjugate scaling fields.  It follows, again by a
linked cluster expansion, that they contribute
\begin{multline} \label{LinearContribution}
\delta f_{imp} = - \frac{1}{2\beta} \sum_{i=c,s,f}  \int_{-\beta/2}^{\beta/2}
d\tau
\lambda^{\prime}_i \cf{{\tilde T}^{(i)}(\tau, 0)}_T \\ + \mbox{subleading terms}
\end{multline}
to the impurity free energy. Note that in contrast to the contribution from
the leading boundary operators in Eq.~(\ref{LinkedCluster}), $\delta f_{imp}$ is
linear in the scaling fields $\lambda^{\prime}_i$.  This reflects the fact
that the energy-momentum tensors are Virasoro descendants of the identity
operator, with nonvanishing one-point functions \cite{DiFrancesco}
\begin{equation} \label{OnePointFunctions}
\cf{{\tilde T}^{(i)}(\tau, 0)}_T = \frac{d_i}{6} (\frac{\pi}{\beta})^2 \, , \ \
\ \ \ \ \ \
d_c = 1, \ \ d_s = d_f = \frac{3}{2} \, .
\end{equation}
Inserting Eq.~(\ref{OnePointFunctions}) into (\ref{LinearContribution}) and
integrating, it follows that the contribution to the impurity specific heat is
\begin{equation}
\delta C_{imp} = \frac{\pi^2}{3} ( \lambda^{\prime}_c  + \frac{3
\lambda^{\prime}_s}{2} +
\frac{3 \lambda^{\prime}_f}{2}) T \, .
\end{equation}
This is the leading correction in Eq.~(\ref{ImpuritySpecificHeat}) that is linear
in the scaling fields: higher-order Virasoro descendants of the identity
operator generate higher powers in temperature. Similarly, higher-order
Kac-Moody descendants of the ${\vphi}^{(s)}$ and ${\vphi}^{(f)}$ operators
produce
subleading corrections to the impurity specific heat that are quadratic in the
scaling fields.

Returning to Eq.~(\ref{ImpuritySpecificHeat}) it is important to notice that the
leading logarithmic terms are independent of the cutoff procedure. We can make
this explicit by introducing the temperature scales $T_s$ and $T_f$ that
characterize the spin and flavor dynamics, and absorb the arbitrary cutoff
$\tau_0$ in the subleading terms:
\begin{widetext}
\begin{multline} \label{ImpuritySpecificHeat2}
C_{imp}  =  \lambda_s^2 A_s \pi^2 T \ln(\frac{T_s}{T}) + \lambda_f^2 A_f \pi^2 T
\ln(\frac{T_f}{T})
         +  \left(\, \lambda_c^2 A_c \tau_0 + \lambda_s^2 A_s
\ln(\frac{1}{\tau_0 T_s}) +
\lambda_f^2 A_f \ln(\frac{1}{\tau_0 T_f}) \, \right) \pi^2 T \\
         +  \mbox{subleading terms} \, . 
\end{multline}
\end{widetext}
It remains to determine the parameters $\lambda_{s,f}, A_{s,f}$ and $T_{s,f}$
that enter the leading terms in Eq.~(\ref{ImpuritySpecificHeat2}). Starting with
the normalization constants of the two-point functions [cf.~Eq.~(\ref{TwoPoint})],
these are defined by \cite{DiFrancesco} $A_{s,f} \equiv \cf{\vJ^{(s,f)}_{-1}
 \cdot {\vphi}^{(s,f)} | \vJ^{(s,f)}_{-1} \cdot {\vphi}^{(s,f)}}$.  Dropping the
indices $(s,f)$, we have that $\cf{\vJ_{-1} \cdot {\vphi} | \vJ_{-1} \cdot
{\vphi}} = \bra{\phi^a} J^a_{1}J^b_{-1} \ket{\phi^b} = \bra{\phi^a} [J^a_{1} ,
J^b_{-1}] + J^b_{-1} J^a_{1} \ket{\phi^b} = \bra{\phi^a} [J^a_{1} , J^b_{-1}]
\ket{\phi^b}$, where the last identity follows from $\ket{\phi^b}$ being a
Kac-Moody primary state.  Using the SU(2)$_2$ Kac-Moody algebra $[J^a_{1} ,
J^b_{-1}] = i\epsilon^{abc}J_0^c + \delta^{ab}$ (common for spin and flavor
sectors), together with the fact that ${\vphi}$ transforms as a spin-1 object
under global SU(2) (generated by $\vJ_{0}$), one straightforwardly arrives at
the result $A_{s,f} = 9$.

Turning to the scaling fields $\lambda_{s,f}$, on dimensional grounds these
must take the form
\begin{equation} \label{ScalingFields}
\lambda_{s,f} = \frac{B_{s,f}}{\sqrt{T_{s,f}}},
\end{equation}
with $B_{s,f}$ dimensionless constants. The scaling fields and 
temperature scales $T_s$ and $T_f$ (which also enter the logarithms explicitly
in Eq.~(\ref{ImpuritySpecificHeat2})) will be extracted from the numerical
solution of the Thermodynamic Bethe Ansatz (TBA).\cite{BolechAndrei} But first we need to discuss the
effect of applied external fields.

\subsection{Magnetic Susceptibility}

In the presence of a magnetic field $B$, applied in the $\hat{z}$ direction,
the term
\begin{displaymath}      \label{MagneticScaling1}
 \int_{-\infty}^{\infty} dx J^z_s (x)
\end{displaymath}
must be added to the scaling Hamiltonian in Eq.~(\ref{ScalingHamiltonian}). Here
$J^z_s = {\small (1/2)} \no{\psidop{\alpha \mu} \sigma^z_{\mu \nu}
\psiop{\alpha \nu}}$ is the $z$ component of the electron spin current that
couples to the field. Passing to the finite-$T$ geometry $\Gamma_c$, this adds a
term
\begin{displaymath}    \label{MagneticScaling2}
B \int_{-\infty}^{\infty} dx \int_{-\frac{\beta}{2}}^{\frac{\beta}{2}} d\tau
\tilde{J}^z_s (\tau, x)
\end{displaymath}
to the impurity effective action in Eq.~(\ref{Partitionfunction}). By a linked
cluster expansion to $O(B^2)$, the impurity magnetic susceptibility
\begin{equation}         \label{MagneticSusceptibility}
\chi_{imp} = - \frac{\del^2 f_{imp}}{\del B^2} \mid_{B = 0}
\end{equation}
gets expressed as
\begin{widetext}
\begin{multline}     \label{MagneticLinkedCluster}
\chi_{imp} =  - \frac{1}{\beta} \sum_{i=c,s,f} \lambda_i \int \! \!
\int_{-\infty}^{\infty}
dx_1 dx_2 \int \, ... \, \int_{-\frac{\beta}{2}}^{\frac{\beta}{2}} d\tau_1 d\tau_2
d\tau_3 \, \cf{\tilde{J}^z_1 \tilde{J}^z_2 
\tilde{{\cal O}}^{(i)}_3}_{T,c} \\
                  + \frac{1}{2\beta} \sum_{i=c,s,f} \lambda_i \lambda_j \int \!
\!
\int_{-\infty}^{\infty} dx_1 dx_2 \int \, ... \,
\int_{-\frac{\beta}{2}}^{\frac{\beta}{2}} d\tau_1 ...
d\tau_4 \, \cf{ \tilde{J}^z_1 \tilde{J}^z_2
\tilde{{\cal O}}^{(i)}_3 \tilde{{\cal O}}^{(j)}_4}_{T, c}  
  + \mbox{subleading terms} \, ,
\end{multline}
\end{widetext}
where ${\cal O}^{(c)} = J, J^z \equiv J^z_s, J^z_i \equiv J^z(\tau_i, x_i),
{\cal O}^{(i)}_k \equiv {\cal O}^{(i)}(\tau_k, 0)$ and ${\cal O}^{(s,f)} =
\vJ^{(s,f)}_{-1} \cdot {\vphi}^{(s,f)}$.  Note that there is no term to zeroth
order in $\lambda_i$, for the same reason that zeroth order terms were absent
in the expansion of the impurity specific heat. Terms linear in $\lambda_c$
and $\lambda_f$ are easily seen to vanish in (\ref{MagneticLinkedCluster}) due
to charge-, spin-, and flavor decomposition of correlation functions:
$\cf{{\cal O}^{(c)}}$ and $\cf{{\cal O}^{(f)}}$ factor out, and with these
operators being Virasoro primaries, the expectation values vanish.  Writing out
the remaining connected three-point function in the spin sector (going back to
the complex plane for notational simplicity), we have
\begin{widetext}
\begin{equation}       \label{ThreePointDecomp}
\cf{J^z_1 J^z_2 {\cal O}^{(s)}_3}_c  = \cf{J^z_1 J^z_2 {\cal O}^{(s)}_3} -
\cf{J^z_1 J^z_2}
\cf{{\cal O}^{(s)}_3} - \cf{J^z_1{\cal O}^{(s)}_3} \cf{J^z_2}
    - \cf{J^z_2{\cal O}^{(s)}_3} \cf{J^z_1}  - 
\cf{J^z_1} \cf{J^z_2} \cf{{\cal O}^{(s)}_3}  \, ,
\end{equation}
with $J_k^z \equiv J^z(z_k)$ and ${\cal O}^{(s)}_k = {\cal O}^{(s)}(z_k)$.
Again, all terms in Eq.~(\ref{ThreePointDecomp}) containing one-point functions
are zero since $J^z$ and ${\cal O}^{(s)}$ are Virasoro primaries. Then,
inserting the operator product expansions \cite{KnizhnikZamolodchikov}
\begin{eqnarray}   
J^z(z_1) {\cal O}^{(s)}(z_2) & = & \frac{3}{(z_1 - z_2)^2} \phi^{(s) z} (z_2) 
+ \mbox{regular terms in} \, (z_1 - z_2) ,       \label{OPE1}   \\
J^z(z_1) \phi^{(s) z} (z_2) & = & \mbox{regular terms in} \, (z_1 - z_2) ,
\label{OPE2}
\end{eqnarray}
\end{widetext}
into $\cf{J^z_1 J^z_2 {\cal O}^{(s)}_3}$, this three-point function collapses
\cite{Details} to a sum of vanishing one-point functions in the neighborhood
of $z_1 \sim z_2 \sim z_3$.  Since the three-point functions in
Eq.~(\ref{ThreePointDecomp}), $\cf{J^z_1 J^z_2 {\cal O}^{(s)}_3}_c$ and $\cf{J^z_1
J^z_2 {\cal O}^{(s)}_3}$, both vanish for infinite separations of the fields,
analyticity then implies that $\cf{J^z_1 J^z_2 {\cal O}^{(s)}_3}_c = 0$ for
any $z_1, z_2, z_3$.  One concludes that there is no contribution to
$\chi_{imp}$ linear in $\lambda_s$.

Turning to the terms in Eq.~(\ref{MagneticLinkedCluster}) quadratic in the scaling
fields, we have (again going back to the complex plane)
\begin{widetext}
\begin{multline}         \label{FourPoint}
\cf{J^z_1 J^z_2 {\cal O}^{(i)}_3 {\cal O}^{(i)}_4}_c = 
\cf{J^z_1 J^z_2 {\cal O}^{(i)}_3 {\cal
O}^{(i)}_4} - \cf{J^z_1 J^z_2} \cf{{\cal O}^{(i)}_3 {\cal O}^{(i)}_4}
   -  \cf{J^z_1 {\cal O}^{(i)}_3 } \cf{J^z_2 {\cal O}^{(i)}_4} -
              \cf{J^z_1 {\cal O}^{(i)}_4 } \cf{J^z_2 {\cal O}^{(i)}_3} -\\
   -  \mbox{terms containing vanishing one-point functions} \, .
\end{multline}
\end{widetext}
As for the case above and as expected on physical grounds, only terms
containing the spin boundary operator ${\cal O}^{(s)}$ can potentially
contribute in
Eq.~(\ref{FourPoint}). Injecting Eq.~(\ref{OPE1}) into $\cf{J^z_1 {\cal O}^{(s)}_3 }
\cf{J^z_2 {\cal O}^{(s)}_4}$ and $\cf{J^z_1 {\cal O}^{(s)}_4 } \cf{J^z_2 {\cal
O}^{(s)}_3}$, these terms are seen to vanish, and one is left with
\begin{equation}    \label{WhatRemains}
\cf{J^z_1 J^z_2 {\cal O}^{(s)}_3 {\cal O}^{(s)}_4} - \cf{J^z_1 J^z_2}
\cf{{\cal O}^{(s)}_3 
{\cal O}^{(s)}_4}\, .
\end{equation}

The four-point function in Eq.~(\ref{WhatRemains}) is easily calculated, again by
exploiting the operator product expansions in Eqs.~(\ref{OPE1}) and (\ref{OPE2}).
Transforming to the finite-$T$ geometry $\Gamma_c$, one finds:\cite{Details}
\begin{widetext}
\begin{multline}   \label{ConnectedFourPoint}
\cf{ \tilde{J}^z (\tau_1, x_1) \tilde{J}^z (\tau_2, x_2)
\tilde{{\cal O}}^{(s)}(\tau_3, 0) \tilde{{\cal O}}^{(s)}(\tau_4, 0)}_{T, c} = \\
 =  9 \Big[ h(\tau_1\! -\! \tau_3\! +\! ix_1) h(\tau_2\! -\! \tau_4 \!+\! ix_2)
+
h(\tau_2\! -\! \tau_3\! +\! ix_2) h(\tau_1\! -\! \tau_4 \!+ \!ix_1) \Big] 
\mid \frac{\beta}{\pi} \sin
\frac{\pi}{\beta} (\tau_3\! - \!\tau_4) \mid^{-1}\, ,
\end{multline}
\end{widetext}
where $h(z) \equiv [(\beta/\pi) \sin (\pi z/\beta)]^{-2}$.
Inserting Eq.~(\ref{ConnectedFourPoint}) into (\ref{MagneticLinkedCluster}) and
carrying out the integration, one finally obtains, to $O(\lambda_s^2)$ [with all other
terms in Eq.~(\ref{MagneticLinkedCluster}) vanishing to this order]:
\begin{equation}       \label{Susceptibility1}
\chi_{imp} = 18 \lambda_s^2 \ln(\frac{T_s}{T}) + O(\lambda_s^3) \, ,  
\end{equation}
where we have subtracted a temperature-independent nonuniversal constant (what
in practice fixes the scale $T_s$). Using the Bethe Ansatz result for $T_s$
and $\lambda_s$ that we give in the next section, we finally obtain a complete
determination of the asymptotic low-temperature behavior of the
susceptibility.

Notice that for the case of the response to an external field in the flavor
degrees of freedom the calculations proceed in a completely analogous way and
the final result is the same with the obvious replacements $T_s \rightarrow
T_f$ and $\lambda_s \rightarrow \lambda_f$.


\subsection{Numerical Fit}

We can use the numerical solution of the TBA
equations in Ref.~\onlinecite{BolechAndrei} to find the values of the conjugate scaling
fields and temperature scales that parametrize the line of boundary conformal
fixed points. This is, to our knowledge, the first time that this kind of fit
is being explicitly carried out.

We start from the following asymptotic expression for the free energy
from which the above results for the entropy, specific heat, and
susceptibilities follow immediately:
\begin{widetext}
\begin{equation}
F_{\mathrm{imp}}\left[  \left\{  t,h_{s,f}\ll t\right\}  \right]
 = 
F_{\mathrm{imp}}^{0}-T\,S_{\mathrm{imp}}^{0}+\sum_{x=s,f}\frac{(3\pi\lambda_{x})^{2}}
{2}\,T^{2}\ln\frac{T}{T_{x}}\,\left(  1+\frac{2 h_{x}^{2}}{\pi^2T^{2}
}\right)  +O\left(  \left\{  T^{2},h_{s,f}^{3}\right\}  \right)
\end{equation}
with $F_{\mathrm{imp}}^{0}=\mathrm{const}$,
$S_{\mathrm{imp}}^{0}=k_B \ln\sqrt{2}$, and $h_{s,f}$ the external field acting
on
the spin or flavor degree of freedom respectively. In order to carry out the
fit, we define the quantities
\begin{equation}
  \tilde{\chi}_{\mathrm{imp}}^{x}  \equiv\frac{2}{h_{x}^{2}}\left(
    F_{\mathrm{imp}}\left[ \left\{ T,h_{x}\ll T,h_{\bar{x}}=0\right\} \right]
    -F_{\mathrm{imp}}\left[  \left\{  T,h_{s,f}=0\right\}  \right]  \right) 
   =18{\lambda_{x}}^{2}\ln\frac{T}{T_{x}}\,+O\left( \left\{ T^{2}
      ,h_{s,f}^{3}\right\} , \right)
\end{equation}
\end{widetext}
which coincide with minus the field susceptibilities in their low-$T$ behavior
and deviate from them as the temperature increases. We have here used the
notation
$\bar{x} = s \, (f)$ if $x = f \, (s)$. The advantage of these
quantities over the usual susceptibilities is that they incorporate the
knowledge that we have about the form that the leading free energy terms take
and they are numerically simpler to compute.

\begin{figure}[tbh]
\begin{center}
\includegraphics[width=0.45\textwidth]{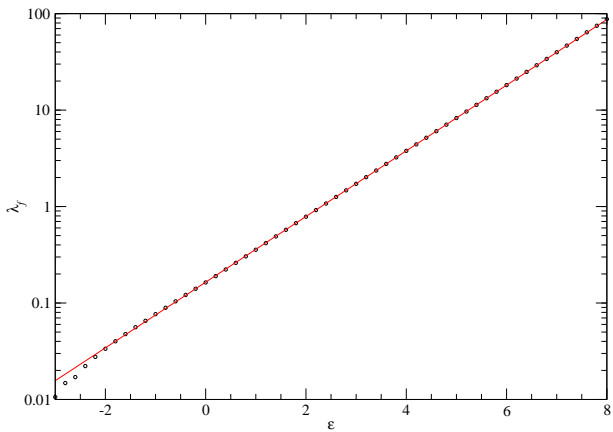}
\caption[Quadrupolar scaling field.]{The dots are the Thermodynamic
Bethe Ansatz numerical results for the flavor scaling field as a
function of $\varepsilon$ (both in units of $\Gamma$) and the line is
our exponential fit.}
\label{lambda}
\end{center}
\end{figure}

We fitted the data from the numerical solution of the TBA equations in the
appropriate temperature range using the form $\tilde{\chi}_{\mathrm{imp}}^{x}
= \tilde{a}_{x}\ln T+\tilde{b}_{x}$ and extracted the desired parameters (as
functions of $\varepsilon$):
$\lambda_{x}=\frac{1}{3}\sqrt{\frac{\tilde{a}_{x}}{2}}$ and $\ln T_{x}%
=-\frac{\tilde{b}_{x}}{\tilde{a}_{x}}$. We also computed $B_{x}=\lambda
_{x}\sqrt{T_{x}}$. In Fig.~\ref{lambda} we show the values obtained for
$\lambda_{f}$ as a function of the doublets energy difference $\varepsilon$
(we have taken $\mu=0$).

\noindent In the same figure we show a fit using the function $\lambda
_{f}={\lambda_0}\ e^{{\gamma}\varepsilon}$ that gives the values:
${\lambda_0} = 0.1640\pm0.0004$ and ${\gamma} = 0.7838\pm0.0004$.
The fit was done in the interval $\left[ -2,8\right] $ and the correlation
coefficient was $\left\vert R\right\vert =0.9999935$. The errors indicated are
statistical ones and do not take into account other sources of error; the
total errors are somewhat larger (as we discuss below).

\begin{figure}[tbh]
\begin{center}
\includegraphics[width=0.45\textwidth]{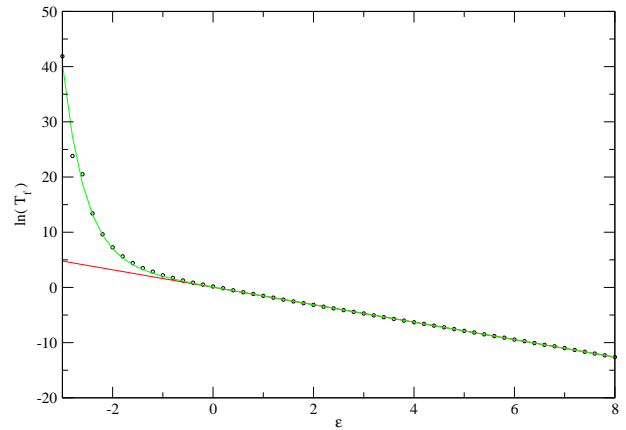}
\caption[Quadrupolar temperature scale.]{The dots are the
Thermodynamic Bethe Ansatz numerical results for the flavor
temperature scale as a function of $\varepsilon$ (both in units of
$\Gamma$) and the lines are two different fits.}
\label{lnTf}
\end{center}
\end{figure}

In Fig.~\ref{lnTf} we show the values obtained this time for $T_{f}$ as a
function of the doublets energy difference.
\noindent We also show in the same figure the results of fitting 
$\ln T_{f}=a+b\ \varepsilon$ which gives the values: $a =
0.033\pm0.012$ and $b =-1.581\pm0.003$. The fit was done for
$\varepsilon>0$ and the correlation coefficient was $\left\vert R\right\vert
=0.9999501$. We next carried out the fit of a more complicated function in
order to capture the behavior for $\varepsilon<0$ as well. We fit using $\ln
T_{f}=a+b\ \varepsilon+c\ e^{d\varepsilon}$ where $a$ and
$b$ are taken to be those determined above. The result of this fit is also
shown in the same figure, and the values obtained for the parameters were
$c = 0.043$ and $d =-2.24$. The fitting was done this time in the
full range of $\varepsilon$, with correlation coefficient $\left\vert
  R\right\vert = 0.998274$.

In the $\varepsilon>0$ region, we combine the results of the two fits to
obtain
\[
B_{f}=\lambda_{f}\sqrt{T_{f}}\approx0.1666\approx1/6 . 
\]

From the expression $T_{l}=\frac{4\Gamma}{\pi^{2}}e^{-\pi/J}$ that was derived
in the Bethe Ansatz analysis,\cite{BolechAndrei} we expect that $b=-\pi/2$ in
the region where $\varepsilon \gg \Gamma$. This value is in fair agreement
with the result of our fit carried out over a larger region ($\varepsilon>0$).
The agreement not only validates our results, but also serves as an indicator
of the magnitude of the total errors of the numerical procedure (always
necessarily larger than the statistical errors of the fit alone). The reader
should notice that the parameter $a$ cannot be compared in a similar way since
it is not a universal quantity and we did not adopt the same
prescriptions in defining $T_{l}$ and $T_{f}$ (the prescription adopted in the
latter case was that there are no constant terms in the definition of
$\tilde{\chi }_{\mathrm{imp}}^{x}$ and that the subleading terms are regular,
i.e.~vanishing, functions of the temperature).

Notice that for negative $\varepsilon$ the scale $T_{f}$ grows exponentially
instead of the linear growth of the scale associated with the quenching of
charge fluctuations. This is no contradiction since the two scales are not
expected to be equal on physical grounds. In other words, there is no reason
to expect that the leading irrelevant operator in the flavor sector will
govern the charge fluctuations when $\varepsilon<0$. One would rather expect
that many operators including those in the charge sector of the theory would
be involved, and in any case the quenching of charge fluctuations takes place
at temperatures far removed from the domain of validity of the scaling
Hamiltonian derived from BCFT. The exponential growth of $T_{f}$ that we
find is indicative of a very rapid suppression of the `importance' of this
operator relative to its counterpart in the spin sector of the theory as
$\varepsilon$ grows negative.

Finally, we also studied the effects of applying a magnetic field and
confirmed numerically that
\[
\left\{
\begin{array}
[c]{c}
\lambda_{s}\left(  \varepsilon-\mu\right)  =\lambda_{f}\left(  \mu
-\,\varepsilon\right)  \\
T_{s}\left(  \varepsilon-\mu\right)  =T_{f}\left(  \mu-\,\varepsilon\right)
\end{array}
\right.
\]
With this observation the parameters of the asymptotic form for the free
energy are now fully determined for any value of $\varepsilon$.

\medskip

A note on some of the details of the numerical calculation is in
order. The results for the free energy in the presence of an applied
field were obtained using a constant field $h_{s,f}=10^{-10}\Gamma$,
which remained always at least two orders of magnitude smaller than the
lower limit of the temperature range used in the fitting of
$\tilde{\chi}_{\mathrm{imp}}^{s,f}$. This was to ensure that we were
always in the basin of attraction of the non-Fermi liquid fixed point
and did not yet start \textit{flowing} towards a Fermi liquid one
driven by the applied field. Since we had to simultaneously capture
scales that were many orders of magnitude apart ($h_{s,f}$
vs $\Gamma$) the numerical calculations had to be done using
``quadruple precision'' arithmetics and the algorithms had to be
designed carefully to assure the precision required. Let us also point
out that when fitting for instance $\tilde{\chi}_{\mathrm{imp}}^{f}$
for negative values of $\varepsilon$ (i.e. determining the temperature
asymptotics of the {\em flavor} linear susceptibility as the system
goes into the overscreened {\em magnetic} moment regime), the
quantities involved become very small and the relative errors grow
rapidly.

\medskip

\begin{figure}[tbh]
\begin{center}
\includegraphics[width=0.4\textwidth]{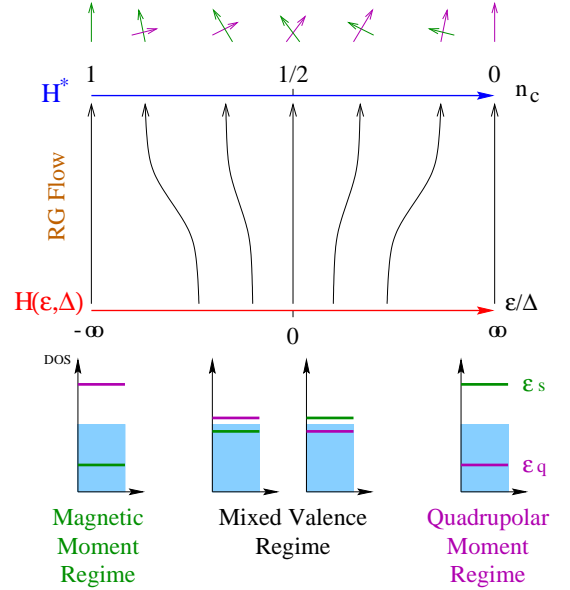}
\caption[Schematic RG Flow.]{Schematic renormalization group flow diagram of
  the two-channel Anderson impurity model. The flow connects the line of
  microscopic theories characterized by the ratio $\varepsilon/\Delta$ (in
  this figure $\Delta\equiv\Gamma$) with
  the line of infrared fixed points parametrized by $n_c$.}
\label{Aflow}
\end{center}
\end{figure}

Our results show that the infrared physics of the two-channel Anderson
impurity is described by a line of non-Fermi liquid fixed points
characterized by a residual entropy $S_{\mathrm{imp}}^{0} =
k_{\mathrm{B}}\ln\sqrt{2}$ and logarithmic temperature divergences in
the specific heat coefficient (i.e.  $\gamma = C_{\mathrm{imp}}/T$)
and the susceptibilities $-$ all these features being already present
in the limiting case embodied by the two-channel Kondo impurity. What
distinguishes the different points on the line are, for instance, the
changing values of the impurity charge valence $n_{c}\left(
\varepsilon-\mu\right)$ (see Fig.~\ref{Aflow}), as well as those of
the scaling fields and the characteristic scales.

Since the specific heat coefficient and the susceptibilities have the same
singular low-temperature divergences along the whole line of fixed points, it
is possible to define the following two \textit{Wilson ratios}:\cite{Wilson}
\begin{equation}
R_{\mathrm{w}}^{s,f}=\lim_{T\rightarrow0}\frac{\chi^{s,f}_{\mathrm{imp}}
/\chi^{s,f}_{\mathrm{bulk}}}{C_{\mathrm{imp}}/C_
{\mathrm{bulk}}}
\end{equation}
where $s \,(f)$ label the {\em magnetic (quadrupolar)} susceptibilities.
This in particular gives for the ''flavor (or quadrupolar) Wilson ratio'':
\begin{align}
R_{\mathrm{w}}^{f}\left(  \varepsilon-\mu\right)  
=\frac{\pi^{2}}{3}\lim_{T\rightarrow0}
\frac{\chi^{f}_{\mathrm{imp}}}{C_{\mathrm{imp}}/T} 
&=\frac{8/3}{1+\left(\lambda_{s}/\lambda_{f}\right)^{2}} \nonumber \\
&=\frac{8/3}{1+e^{-4{\gamma}(\varepsilon-\mu)}}
\end{align}
with ${\gamma} = 0.7838\pm0.0004$. These ratios 
are another characteristic that varies along
the line of fixed points. In particular, we have that
$R_{\mathrm{w}}^{f}\left( +\infty\right) = 8/3$ in agreement with the
result for the two-channel \textit{quadrupolar} Kondo model and, as
expected, $R_{\mathrm{w}} ^{f}\left( -\infty\right) =0$, since this is
the \textit{magnetic} Kondo limit. By symmetry arguments, we find that
$R_{\mathrm{w}}^{s}\left( \varepsilon-\mu\right) =R_{\mathrm{w}}
^{f}\left( \mu-\varepsilon\right)$ and $R_{\mathrm{w}}^{f}\left(
0\right) = 4/3$. Finally, a simple calculation yields the universal
relation: $R_{\mathrm{w}}^{f}\left(\varepsilon-\mu\right) +
R_{\mathrm{w}}^{s}\left(\varepsilon-\mu\right) = 8/3$.

The measurement of $R_{\mathrm{w}}^{s,f}$ is a very
sensitive way to experimentally determine which doublet is the lower one in
energy, whether the impurity ion ground state is magnetic or
quadrupolar. This can be very useful if the system is in the mixed valence
region $\left\vert \varepsilon-\mu\right\vert <\Gamma$, as was thought to be
the case in \textrm{UBe}$_{13}$. This question  triggered the
interest in measuring the nonlinear susceptibility of this
compound \cite{Ramirez} and its answer is still a source of some
controversy.\cite{Aliev1}


\subsection{Nonlinear Magnetic Susceptibility}

The nonlinear magnetic susceptibility $\chi_3$ measures the leading
nonlinearity in the magnetization
\begin{equation}        \label{Magnetization}
M = \chi B + \frac{1}{6} \chi_3 B^3 + ...
\end{equation}
in the direction of an applied field $B$. As follows from the analysis in
Ref.~\onlinecite{Lea}, a quadrupolar moment contributes nontrivially to $\chi_3$ via
its
coupling to the square of the magnetic field.  In our formalism this coupling
adds a term $\alpha B^2 J_f^z$ to the scaling Hamiltonian, where $\alpha$ is a
constant and $J_f^z = \small{(1/2)} \no{ \psidop{\alpha \mu} \tau^z_{\alpha
\beta} \psiop{\beta \mu} }$ is the component of the quadrupolar moment
along the field direction, with $\tau^z$ the diagonal SU(2) flavor generator.

The fact that the quadrupolar moment couples to $B^2$ permits the
determination of the quadrupolar susceptibility from measurements of the
(nonlinear) magnetic susceptibility.  Here we shall explore how the leading
contribution to $\chi_3$ from the Anderson impurity, $\chi_{3, imp}$ call it,
behaves as one goes along the critical line from the quadrupolar to the
magnetic two-channel Kondo fixed points.

We thus have to consider the extended scaling Hamiltonian
\begin{widetext}
\begin{eqnarray}   \label{ExtendedScaling}
H_{scaling} & = & H^{\ast} + \lambda_s {\cal O}^{(s)}(0) + \lambda_f {\cal
O}^{(f)}(0)
+ B \int_{-\infty}^{\infty} dx J_s^z(x) + \alpha B^2 \int_{-\infty}^{\infty} dx
J_f^z(x)
 + \, \mbox{subleading terms}\, , 
\end{eqnarray}
\end{widetext}
where we have removed ``by hand'' terms from the charge sector (which do not
contribute to $\chi_{3, imp}$).  Given
\begin{equation}       \label{NLS3}
\chi_{3, imp} = - \frac{\partial^4 f_{imp}}{\partial B^4} \mid_{B=0}    
\end{equation}
we now have to carry the linked cluster expansion of the free energy to
$O(B^4)$, in analogy to the $O(B^2)$ expansion in the calculation of the
linear susceptibility in Sec.~(IV.C). To second order in the scaling fields
$\lambda_s$ and $\lambda_f$ we then find that $f_{imp}$ can receive possible
contributions from 15 different terms generated by $H_{scaling}$ in
Eq.~(\ref{ExtendedScaling}). These terms have the structure
\begin{widetext}
\begin{displaymath}
\alpha^{\ell_0} B^4 \lambda_s^{k_0} \lambda_f^{\ell_0} \cf{\Pi_{i=0,1,...i_0}
J_s^z(z_i) 
\Pi_{j=0,1,...,j_0} J_f^z(z_{j+i_0})
\Pi_{k=0,1,k_0} {\cal O}_s(z_{k+j_0+i_0}) \Pi_{\ell =0,1,\ell_0} 
{\cal O}_f(z_{\ell +k_0+j_0+i_0})} \, , 
\end{displaymath}
where $k_0+\ell_0 = 1$ or $2$, and $i_0+2j_0 = 4$ (with $i_0=0, 2, 4; \, j_0 =
0, 2$),
and where 
$\Pi_{m=0} \, ... \equiv 1 \ (m = i, j, k, \ell)$. By repeated use of the
operator product expansions in Eqs.~(\ref{OPE1}) and (\ref{OPE2}), a
straightforward analysis reveals that among these fifteen terms only one is
nonvanishing.\cite{FOOTNOTE2} Passing to the finite-$T$ geometry $\Gamma_c$, it
has the form
\begin{multline}   
\cf{ \tilde{J}^z_f (\tau_1, x_1) \tilde{J}^z_f (\tau_2, x_2)
\tilde{{\cal O}}^{(f)}(\tau_3, 0) \tilde{{\cal O}}^{(f)}(\tau_4, 0)}_{T, c} \\
 =  9 \Big[ h(\tau_1 - \tau_3 + ix_1) h(\tau_2 - \tau_4 + ix_2) +
h(\tau_2 - \tau_3 + ix_2) h(\tau_1 - \tau_4 + ix_1) \Big] \mid \frac{\beta}{\pi}
\sin
\frac{\pi}{\beta} (\tau_3 - \tau_4) \mid^{-1}\, , 
\end{multline}
\end{widetext}
which is the same connected four-point function that we encountered in
Eq.~(\ref{ConnectedFourPoint}), but with flavor now taking the place of spin.
By comparison with the analysis that follows in Sec.~(IV.C) we immediately
conclude that the leading contribution to the nonlinear susceptibility has the
same logarithmic form as that for the linear susceptibility, but now
parametrized by the ``quadrupolar'' scaling field $\lambda_f$ and temperature
scale $T_f$:
\begin{equation}       \label{NLSusceptibility1}
\chi_{3,imp} = 18 \lambda_f^2\alpha^2\ln(\frac{T_f}{T}) + O(\lambda_f^3) \, .
\end{equation}
Combined with the Bethe Ansatz fit of $\lambda_f$ and $T_f$ from the
previous section, Eq.~(\ref{NLSusceptibility1}) completely specifies
the leading scaling term of the nonlinear magnetic susceptibility.

\section{Summary}
\label{section5}

We have carried out in this paper a detailed analysis of the low-temperature
thermodynamics of the two-channel Anderson impurity model, identifying and
fitting a
boundary conformal field theory to the Bethe Ansatz solution of the
model.  Combining the methods allowed us to completely determine the various
parameters and scales introduced in the BCFT and, in terms of these, give a
complete description of the line of fixed points characterizing the low-energy
physics. 

As we mentioned in the Introduction, having identified the scaling operators
the asymptotic dynamical properties can in principle be calculated. This
allows the study of resistivities, optical conductivities, and Green's
functions in general. These results will be presented in a subsequent work.

\begin{acknowledgments}

This work was started while one of us (H.~J.) was visiting Rutgers
University. He thanks the Physics Department for its hospitality, and
the Swedish Research Council and the STINT foundation for financial
support. We are grateful to P.~Coleman and P.~Fonseca for useful and
enlightening discussions.

\end{acknowledgments}



\begin{references}
\bibitem{FOOTNOTE} Note that while the lattice physics underlying the
Kondo model is not expected to respect charge conjugation symmetry,
the Kondo model itself is invariant under the corresponding
transformation $\psiop{\sigma}(x) \rightarrow
\epsilon^{\phdagger}_{\sigma \mu} \psidop{\mu}(x)$. The reason for
this is that the normal ordering of the electron fields in the model
subtracts the symmetry-breaking terms. We may remove this accidental
symmetry by explicitly introducing operators that break charge
conjugation, of which the leading one is the charge current
$J(0)$. However, in the case of the Kondo model there is no tunable
parameter that can change its conjugate scaling field $\lambda_c$
[in contrast to the Anderson model considered here where $\lambda_c =
\lambda_c(\varepsilon_s, \varepsilon_q) = - n_c(\varepsilon_s, \varepsilon_q)/4$
(cf.~Eq.~(\ref{PotentialScattering})].  For this reason the marginal charge
current plays no role in the Kondo model other than generating trivial
subleading corrections to the leading impurity critical behavior.
\cite{AffleckLudwig1}
\bibitem{FOOTNOTE2} By inspection one also verifies that there is no
  contribution to $O(B^2$) coming from the coupling of the magnetic field to
  the quadrupolar moment. Hence, the result derived in Sec.~(IV.C) for the
  {\em linear} susceptibility remains valid also when this coupling is
  included in the scaling Hamiltonian.
\end{references}
\end{document}